\documentclass[aps,prb,twocolumn,showpacs,superscriptaddress]{revtex4-1}

\usepackage{graphicx}
\usepackage{amsmath}
\usepackage{bm}
\usepackage{longtable}
\usepackage{xspace}
\usepackage{color}

\newcommand{\lmo}{LaMn$_7$O$_{12}$}
\newcommand{\nmo}{NdMn$_7$O$_{12}$}

\newcommand{\cmo}{CeMn$_7$O$_{12}$}
\newcommand{\smo}{SmMn$_7$O$_{12}$}
\newcommand{\emo}{EuMn$_7$O$_{12}$}
\newcommand{\ymo}{YMn$_7$O$_{12}$}

\begin{document}

\title{The magnetic structures of rare-earth quadruple perovskite manganites \textit{R}Mn$_7$O$_{12}$}

\author{R. D. Johnson}
\email{roger.johnson@physics.ox.ac.uk}
\affiliation{Clarendon Laboratory, Department of Physics, University of Oxford, Oxford, OX1 3PU, United Kingdom}
\author{D. D. Khalyavin}
\author{P. Manuel}
\affiliation{ISIS facility, Rutherford Appleton Laboratory-STFC, Chilton, Didcot, OX11 0QX, United Kingdom}
\author{L. Zhang}
\author{K. Yamaura}
\affiliation{Research Center for Functional Materials, National Institute for Materials Science (NIMS), Namiki 1-1, Tsukuba, Ibaraki 305-0044, Japan}
\affiliation{Graduate School of Chemical Sciences and Engineering, Hokkaido University, North 10 West 8, Kita-ku, Sapporo, Hokkaido 060-0810, Japan}
\author{A. A. Belik}
\affiliation{Research Center for Functional Materials, National Institute for Materials Science (NIMS), Namiki 1-1, Tsukuba, Ibaraki 305-0044, Japan}

\date{\today}

\begin{abstract}
We report a neutron powder diffraction study of $R$Mn$_7$O$_{12}$ quadruple perovskite manganites with $R$ = La, Ce, Nd, Sm, and Eu. We show that in all measured compounds concomitant magnetic ordering of the $A$ and $B$ manganese sublattices occurs on cooling below the N$\mathrm{\acute{e}}$el temperature. The respective magnetic structures are collinear, with one uncompensated Mn$^{3+}$ moment per formula unit as observed in bulk magnetisation measurements. We show that both \lmo\ and \nmo\ undergo a second magnetic phase transition at low temperature, which introduces a canting of the $B$ site sublattice moments that is commensurate in \lmo\ and incommensurate in \nmo. This spin canting is consistent with a magnetic instability originating in the $B$ site orbital order. Furthermore, \nmo\ displays a third magnetic phase transition at which long range ordering of the Nd sublattice modifies the periodicity of the incommensurate spin canting. Our results demonstrate a rich interplay between transition metal magnetism, orbital order, and the crystal lattice, which may be fine tuned by cation substitution and rare earth magnetism.
\end{abstract}

\pacs{}

\maketitle

\section{Introduction}

The quadruple perovskite manganites with chemical formula $AA'_3$Mn$_4$O$_{12}$ derive from the simple perovskite lattice, and a pattern of large octahedral tilts ($a^+a^+a^+$ in Glazer notation) accommodates a 1:3 ordering of $A$ and $A'$ cations. This chemical order expands the range of manganese-based oxides in which to study the interplay between charge, orbital, and spin degrees of freedom \cite{Vasiliev07,Gilioli14}. Furthermore, topical physical properties such as low-field magnetoresistance\cite{Zeng99} and multiferroic behaviour \cite{Mezzadri09_1,Johnson12} have been found in the quadruple perovskite manganites.

When $A'$ = Mn$^{3+}$ a family of manganites is formed that nominally contain Mn$^{4+}$ and Jahn-Teller active Mn$^{3+}$ cations in different proportions on the $B$ sites. The Mn$^{4+}$:Mn$^{3+}$ ratio is determined by the oxidation state of the $A$ cations, which may be monovalent Na$^+$, divalent Mn$^{2+}$, Cd$^{2+}$, Ca$^{2+}$, Sr$^{2+}$, Pb$^{2+}$, or trivalent $R^{3+}$ ($R$ = rare earth including Y and Bi)\cite{Marezio73,BOCHU74,Prodi09,Mezzadri09_1,Mezzadri09_2,Johnson12,Locherer12,Ovsyannikov13,Glazkova15,Belik16,Verseils17}. Despite the successful synthesis of a large number of quadruple perovskites, the only trivalent $A$ site manganites whose physical properties had been reported prior to our work were \lmo,\cite{Prodi09} PrMn$_7$O$_{12}$, \cite{Mezzadri09_2}, BiMn$_7$O$_{12}$,\cite{Imamura08,Mezzadri09_1} and YMn$_7$O$_{12}$\cite{Verseils17}.

At high temperature \lmo\ crystallises in the cubic, $Im\bar{3}$ quadruple perovskite structure \cite{Okamoto09}. As for all $R^{3+}$Mn$_7$O$_{12}$ compounds every $B$ site is occupied by a Jahn-Teller active Mn$^{3+}$ ion, and below 650 K the crystal symmetry is lowered to monoclinic $I2/m$ as a result of $d_{3z^2-r^2}$ orbital ordering within the $ac$ plane, with nearest neighbour orbitals orientated orthogonal to each other --- the same pattern of orbital ordering supported by the simple perovskite LaMnO$_3$ \cite{Murakami98,Matsumoto70}. This crystal symmetry and orbital order persists down to low temperature. BiMn$_7$O$_{12}$ and YMn$_7$O$_{12}$ adopt the same pattern of orbital order as \lmo, but the former shows complex structural behaviour as a function of temperature \cite{Belik17}, and the latter undergoes a structural phase transition at 200 K that is apparently isostructural \cite{Verseils17}, yet its origin remains unknown. The structural behaviour of PrMn$_7$O$_{12}$ is complicated by the presence of two polymorphs, one with $I2/m$ symmetry, and the other $R\bar{3}$, which are highly sensitive to synthesis conditions. However, the monoclinic phase fraction is understood to be similar to the La and Bi compounds \cite{Mezzadri09_2}. PrMn$_7$O$_{12}$ was omitted from the present study due to these structural complications.

\lmo\ and BiMn$_7$O$_{12}$ are reported to undergo two magnetic phase transitions at $T_1$ = 78 K and $T_2$ = 21 K, and $T_1$ = 59 K and $T_2$ = 28 K, respectively \cite{Prodi09,Mezzadri09_1}. In \lmo\ it was found that the $B$ site manganese moments ordered below $T_1$ with a \textbf{k}=(0,0,0), antiferromagnetic alignment in the $ac$ plane, with ferromagnetic stacking along the $b$ axis. The $A$ site manganese moments were then found to order independently at $T_2$ with propagation vector \textbf{k}=(0,1,0). In \ymo, the same \textbf{k}=(0,0,0) antiferromagnetic ordering of $B$ site moments was reported below $T_1$ = 108 K, however, no lower temperature magnetic phase transition was observed. We note that the $B$ site magnetic structure of both compounds is somewhat surprising, as the established orbital order favours ferromagnetic planes stacked antiferromagnetically, as observed in LaMnO$_3$ \cite{Wollan1955,goodenough55}. We discuss this point in detail later.

In this article we report the synthesis and characterisation of polycrystalline samples of $R$Mn$_7$O$_{12}$ compounds, where $R$ = La, Ce, Nd, Sm, and Eu. We demonstrate that in all measured compounds the $A$ and $B$ manganese sublattices magnetically order concomitantly on coooling below the N$\mathrm{\acute{e}}$el temperature --- contrary to previous reports on \lmo\cite{Prodi09} and \ymo\cite{Verseils17}. The refined magnetic structure is collinear with one uncompensated $A$ site Mn$^{3+}$ magnetic moment per formula unit, as observed in the bulk magnetisation. We show that both \lmo\ and \nmo\ display a second phase transition at low temperature, which is related to a spin canting of the $B$ site sublattice that is commensurate in \lmo, and incommensurate in \nmo. This spin canting is consistent with an underlying magnetic instability originating in the $B$ site orbital order. Furthermore, \nmo\ displays a third magnetic phase transition at which long range ordering of the Nd sublattice modifies the periodicity of the incommensurate spin canting. The article is organised as follows. We first present details of the sample synthesis and experimental techniques in Section \ref{exsec}. A detailed analysis of neutron powder diffraction data measured from \lmo\ and \nmo\ is given in Sections \ref{LaSec} and \ref{NdSec}, followed by a brief comparison with diffraction data collected on \cmo, \smo, and \emo\ in Section \ref{CSESec}. In Section \ref{dissec} we discuss the results with particular focus on orbital order and bulk magnetisation, and finally, conclusions are drawn in Section \ref{consec}.

\section{\label{exsec}Experiment}

Polycrystalline $R$Mn$_7$O$_{12}$ samples with $R$ = La, Nd, Sm, and Eu were prepared from stoichiometric mixtures of Mn$_2$O$_3$ and $R_2$O$_3$ (99.9\%). \cmo\ was prepared from stoichiometric mixtures of Mn$_2$O$_3$, Mn$_3$O$_4$ (99.99\%), and CeO$_2$ (99.99\%). Single-phase Mn$_2$O$_3$ was prepared from commercial MnO$_2$ (99.99\%) by heating in air at 923 K for 24 h. The mixtures were placed in Au or Pt capsules and treated at 6 GPa and about 1570-1670 K for 2 h (heating time to the synthesis temperature was 10 min) in a belt-type high-pressure apparatus. After the heat treatments, the samples were quenched to room temperature, and the pressure was slowly released. All obtained samples were black pellets. 

Variable temperature magnetisation measurements of dense pellets from the growth were performed using a SQUID magnetometer (Quantum Design MPMS) on cooling in an applied magnetic field of 1 T from 350 to 2 K. Isothermal magnetization measurements were performed between -7 and 7 T at different temperatures. Neutron powder diffraction measurements were performed on the WISH time-of-flight diffractometer \cite{Chapon11} at ISIS, the UK Neutron and Muon Spallation Source. Samples were lightly packed into thin 3mm cylindrical vanadium cans to minimse the effects of neutron absorption (large for Sm and Eu), and mounted within a $^4$He cryostat. Data were collected with high counting statistics at a fixed temperature within each magnetic phase, including paramagnetic for reference. Data were also collected with lower counting statistics on warming in the temperature range 1.5 K to 100 K. All diffraction data were refined using Fullprof \cite{Rodriguezcarvaja93}.

\section{Results}

\subsection{\label{LaSec}LaMn$_7$O$_{12}$}

\begin{figure}
\includegraphics[width=8.8cm]{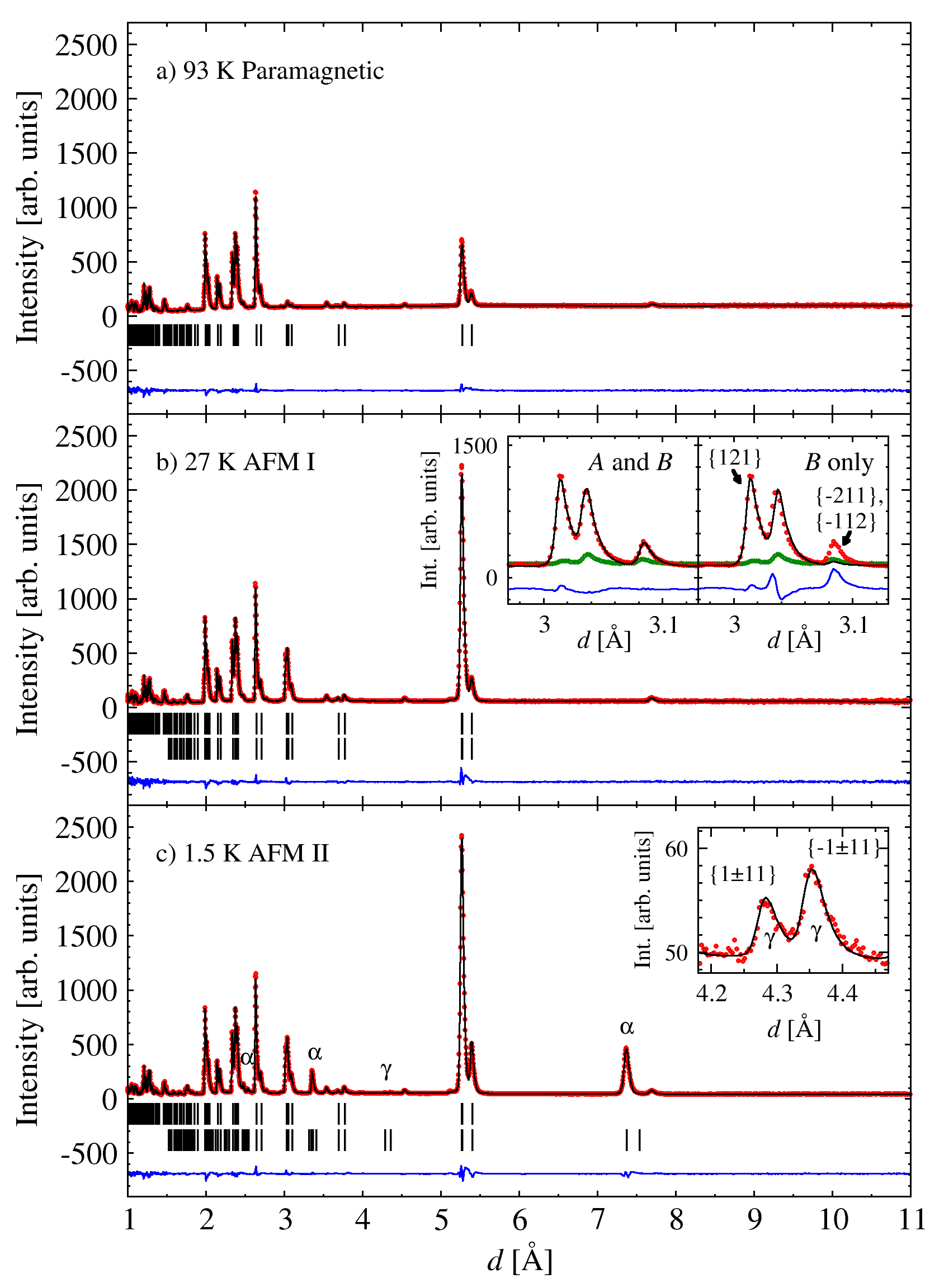}
\caption{\label{FIG:La_NPD_fig}Neutron powder diffraction data measured from \lmo\ in bank 2 (average $2\theta = 58.3^\circ$) of the WISH diffractometer. Diffraction data (red points) collected in the paramagnetic, AFM I, and AFM II phases are shown in panes a), b) and c), respectively. The fitted nuclear (top tick marks) and magnetic (bottom tick marks) structural models are shown as a solid black line. Tick marks corresponding to Mn$_3$O$_4$ (3.4 wt\%), LaMnO$_3$ (0.5 wt\%), and La(CO$_3$)(OH) (0.9 wt\%) impurity phases have been omitted for clarity. A difference pattern ($I_\mathrm{obs}-I_\mathrm{calc}$) is given as a blue solid line at the bottom of each pane.}
\end{figure}

\begin{figure}
\includegraphics[width=8.8cm]{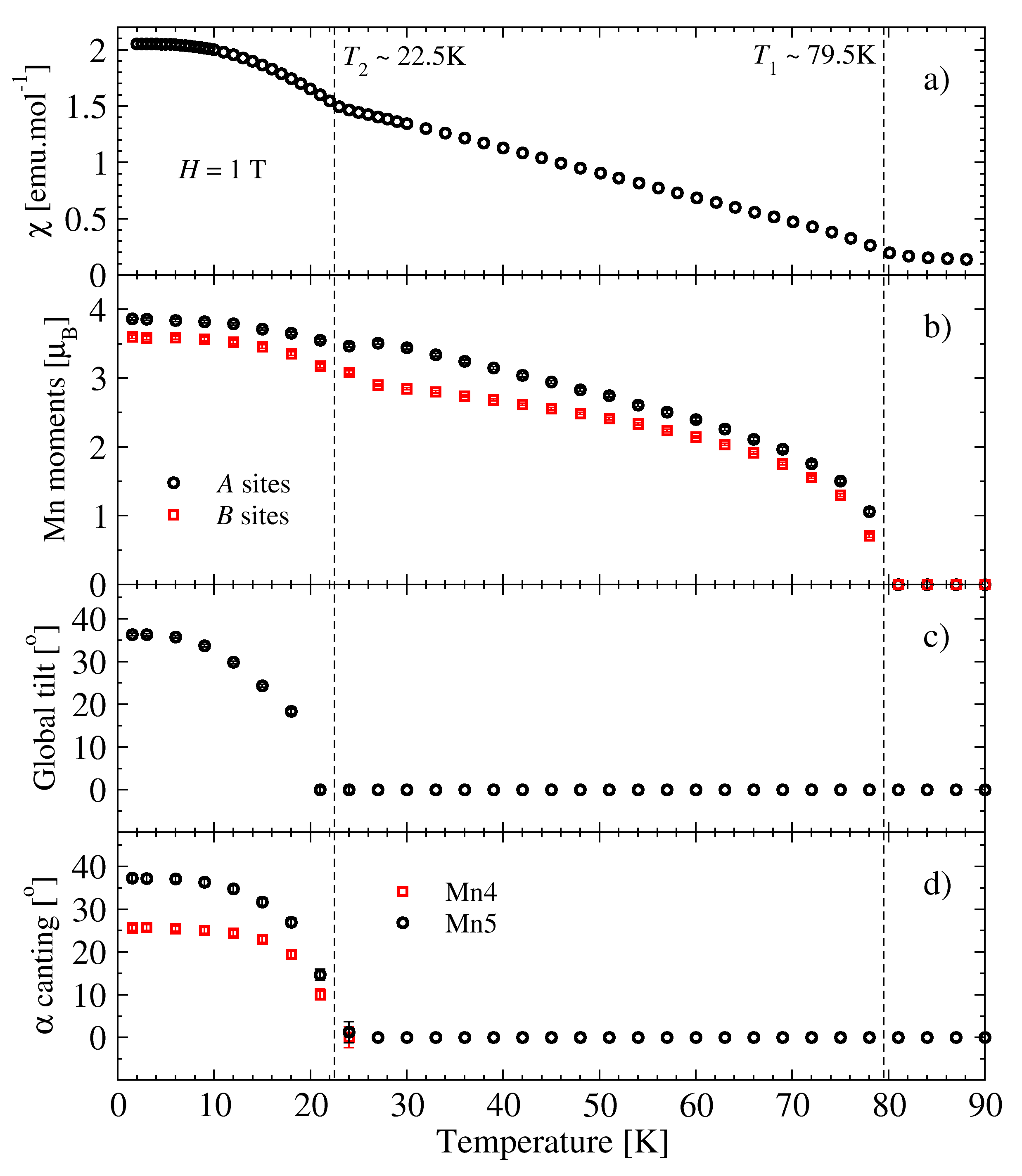}
\caption{\label{FIG:La_temp_dep_fig}The \lmo\ temperature dependence of a) field cooled magnetic susceptibility, b) the $A$ and $B$ site ordered magnetic moments, c) the global tilt of the \textbf{k}=(0,0,0) magnetic structure component, and d) the canting angle of the Mn4 and Mn5 moments due to the evolution of the \textbf{k}=(0,1,0) $\alpha$ mode, defined as $\tan^{-1}(m_\alpha/m_\chi)$.}
\end{figure}

Variable temperature magnetic susceptibility data measured from our polycrystalline \lmo\ sample (Figure \ref{FIG:La_temp_dep_fig}a) showed signatures of two magnetic phase transitions at $T_1 \sim 79.5$ K and $T_2 \sim 22.5$ K, which are both consistent with previous observations\cite{Prodi09}. Neutron powder diffraction data collected at 93 K, within the paramagnetic phase of \lmo, was fit with a refined model based upon the published crystal structure \cite{BOCHU74,Prodi09}, and excellent agreement with the data was achieved (Figure \ref{FIG:La_NPD_fig}a). The sample was found to be 95.2 wt\% pure, which is typical for samples synthesised under high pressure conditions. Structural parameters are summarised in Table \ref{TAB:Crystal}, where atomic displacement parameters have been omitted as in several compounds strong neutron absorption prevented their reliable determination.

\begin{table*}
\caption{\label{TAB:Crystal} Crystal structure parameters ($I2/m$, $Z$ = 2) in the paramagnetic phase of $R$Mn$_7$O$_{12}$, with $R = $La, Ce, Nd, Sm, and Eu refined at 93, 90, 100, 100, and 100 K, respectively. Wyckoff positions are as follows. $A$ sublattice; $R$: 2a, Mn1: 2b, Mn2: 2c, Mn3: 2d. $B$ sublattice; Mn4: 4e, Mn5: 4f. Oxygen; O1 \& O2: 4i, O3 \& O4: 8j. The values of $T_1$ are given at the bottom.}
\begin{ruledtabular}
\begin{tabular}{c | c c c c c  }
 & \lmo & \cmo & \nmo & \smo & \emo \\
\hline
\multicolumn{6}{l}{Lattice parameters}\\
$a$ [$\mathrm{\AA}$] & 7.5219(1) & 7.4966(2) & 7.4885(1) & 7.4854(2) & 7.4769(3) \\
$b$ [$\mathrm{\AA}$] & 7.36243(8) & 7.3511(2) & 7.33241(8) & 7.3287(2) & 7.3246(2) \\
$c$ [$\mathrm{\AA}$] & 7.51685(9) & 7.4915(2) & 7.4850(1) & 7.4837(2) & 7.4748(2) \\
$\beta$ [$^\circ$] & 91.309(1) & 91.232(1) & 91.256(1) & 91.231(2) & 91.215(2) \\
$V$ [$\mathrm{\AA}^3$] & 416.168(9) & 412.75(2) & 410.892(9) & 410.45(2) & 409.27(2) \\
\multicolumn{6}{l}{Fractional coordinates}\\
O1 \hspace*{\fill} $x$ & 0.1676(5) & 0.1671(5) & 0.1666(4) & 0.1651(8) & 0.1632(8) \\
   \hspace*{\fill} $z$ & 0.3084(5) & 0.3079(5) & 0.3056(4) & 0.3055(7) & 0.3034(7) \\
O2 \hspace*{\fill} $x$ & 0.1783(6) & 0.1786(6) & 0.1775(6) & 0.1781(8) & 0.178(1) \\
   \hspace*{\fill} $z$ & 0.6882(7) & 0.6857(7) & 0.6877(6) & 0.685(1) & 0.687(1) \\
O3 \hspace*{\fill} $x$ & 0.0151(4) & 0.0153(4) & 0.0150(4) & 0.0153(6) & 0.0146(7) \\
   \hspace*{\fill} $y$ & 0.3100(4) & 0.3089(4) & 0.3084(4) & 0.3065(6) & 0.3054(6) \\
   \hspace*{\fill} $z$ & 0.1722(4) & 0.1744(4) & 0.1722(4) & 0.1737(6) & 0.1743(6) \\
O4 \hspace*{\fill} $x$ & 0.3125(4) & 0.3112(4) & 0.3111(3) & 0.3107(6) & 0.3097(6) \\
   \hspace*{\fill} $y$ & 0.1758(4) & 0.1755(4) & 0.1753(4) & 0.1749(6) & 0.1739(6) \\
   \hspace*{\fill} $z$ & -0.0137(4) & -0.0123(4) & -0.0126(3) & -0.0123(5) & -0.0124(6)\\
\multicolumn{6}{l}{Fit reliability parameters}\\
R [\%] & 3.3 & 3.65 & 3.19 & 3.4 & 3.3 \\
$w$R [\%] & 3.4 & 4.6 & 3.28 & 3.5 & 3.2 \\
R$_\mathrm{Bragg}$ [\%] & 4.6 & 3.61 & 3.38 & 4.9 & 6.1\\
\multicolumn{6}{l}{N$\mathrm{\acute{e}}$el temperature}\\
$T_1$ [K]  & 79.5 & 80 & 85 & 87 & 87
\end{tabular}
\end{ruledtabular}
\end{table*}

\begin{figure}
\includegraphics[width=8.8cm]{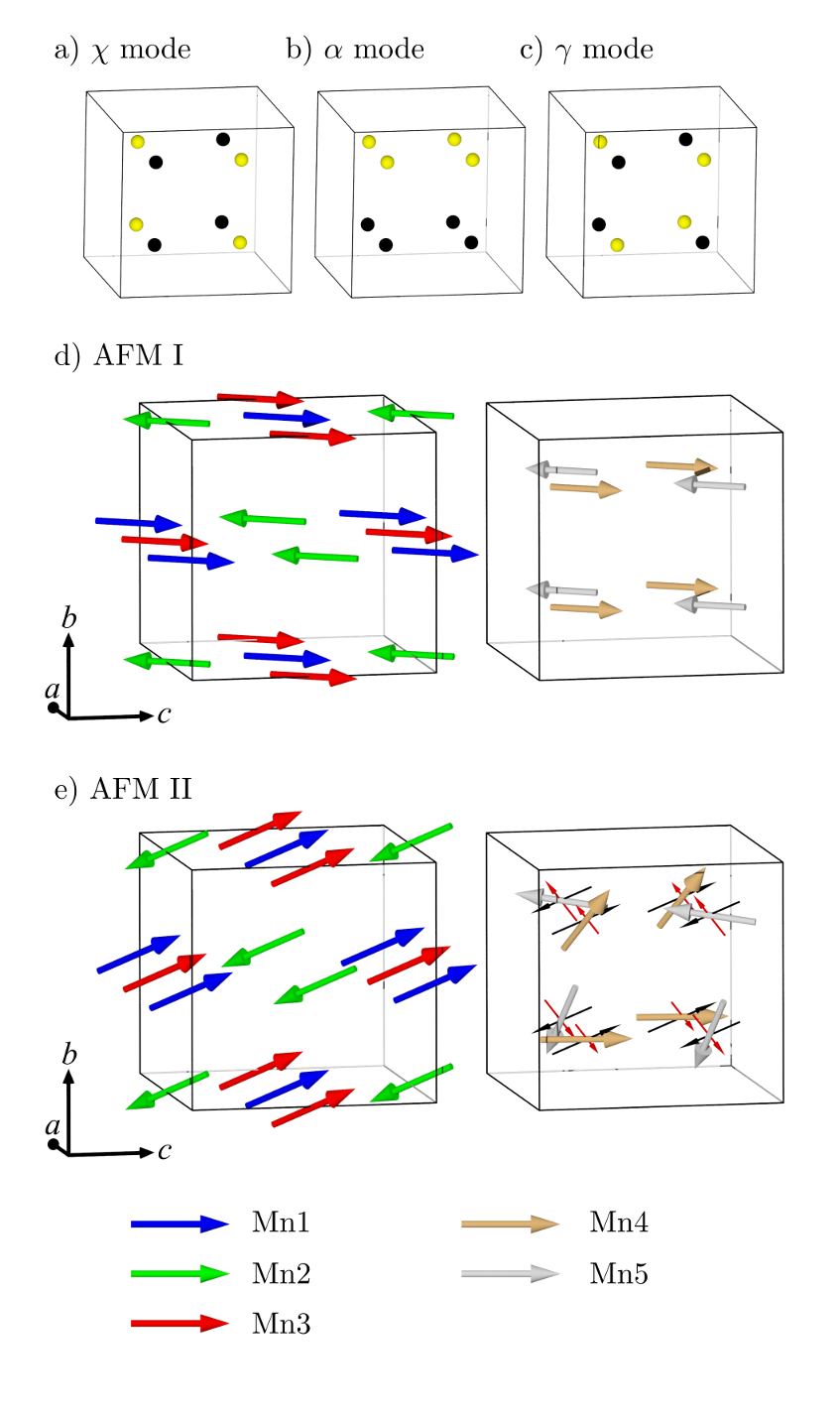}
\caption{\label{FIG:La_mag_struc_fig}(a)-(c) Antiferromagnetic modes of the $B$ site sublattice, where yellow and black spheres represent moments of opposite sign. The refined magnetic structures of the AFM I and AFM II phases of \lmo\ are shown in (d) and (e) respectively, with the $A$ sublattice on the left and the $B$ sublattice on the right. The orthogonal $\chi$ and $(\alpha+\gamma)$ components of the ground state $B$ sublattice magnetic structure are shown as thin black and red arrows, respectively. In all panes the $I2/m$ quadruple perovskite unit cell is drawn in black lines.}
\end{figure}

On cooling below $T_1$ additional magnetic diffraction intensities were observed, which were unambiguously indexed with the propagation vector \textbf{k}=(0,0,0), as reported \cite{Prodi09}. We label this phase AFM I. Symmetry analysis using the \textsc{isotropy} suite\cite{Campbell06,Stokes07} showed that two irreducible representations, $\Gamma_1^+$ and $\Gamma_2^+$, enter into the decomposition of the magnetic $\Gamma$-point reducible representation for the relevant Wyckoff positions (see Table \ref{TAB:irreps} of the Appendix for the basis vectors of $\Gamma_1^+$ and $\Gamma_2^+$). Exhaustive testing of possible magnetic structures found that only one minimal model, which transforms according to $\Gamma_2^+$, faithfully reproduced the neutron diffraction data shown in Figure \ref{FIG:La_NPD_fig}b. The fitted AFM I magnetic structure (Figure \ref{FIG:La_mag_struc_fig}d) is collinear with all magnetic moments lying within the $ac$ plane. Out of plane tilts of $B$ site moments, which are allowed within the symmetry of $\Gamma_2^+$, did not improve the quality of the fit. The $B$ site moments were ordered antiferromagnetically within the $\chi$ mode shown in Figure \ref{FIG:La_mag_struc_fig}a (for clarity we use $\chi$, $\alpha$, and $\gamma$ to denote $B$ site magnetic structure modes typically referred to as C-type, A-type, and G-type, respectively). The $A$ site moments were ordered in a ferrimagnetic structure with two uncompensated moments per unit cell (one moment per formula unit).  The refined parameters of the magnetic structure are given in Table \ref{TAB:Magnetic}. The moment magnitudes on all $A$ sites were constrained to be the same, as were those on all $B$ sites, and their temperature dependence is shown in Figure \ref{FIG:La_temp_dep_fig}b.

\begin{table*}
\caption{\label{TAB:Magnetic} Magnetic structure parameters refined in the AFM I phase of $R$Mn$_7$O$_{12}$. Moment directions are given in spherical coordinates defined such that $m_z = m\cos(\theta)~||~c$, $m_x = m\sin(\theta)~||~a^*$, and $m_y = 0~||~b$. Mn1 and Mn3 moments are parallel, and antiparallel to Mn2 moments. All moments of the Mn4 sublattice are parallel, and antiparallel to those of the Mn5 sublattice.}
\begin{ruledtabular}
\begin{tabular}{l | c c c c c }
 & \lmo & \cmo & \nmo & \smo & \emo \\
\hline
$T$ [K] & 27 & 1.5 & 30 & 1.5 & 1.5 \\
\\
Mn1  \hspace*{\fill} $m [\mu_\mathrm{B}]$ & 3.51(2) & 3.03(3) & 3.28(3) & 3.45(4) & 3.11(6) \\
     \hspace*{\fill} $\theta [^\circ]$ & -32.7(3) & -38.7(3) & -33.1(3) & -34.7(4) & -32.1(5) \\
    \\
Mn4  \hspace*{\fill} $m [\mu_\mathrm{B}]$ & 2.90(2) & 2.97(1) & 3.00(2) & 3.27(3) & 3.24(3) \\
     \hspace*{\fill} $\theta [^\circ]$ & -32.7(3) & -38.7(3) & -33.1(3) & -34.7(4) & -32.1(5) \\
    \\
\multicolumn{6}{l}{Fit reliability parameters}\\
R [\%] & 4.56 & 4.57 & 4.19 & 4.13 & 6.94 \\
$w$R [\%] & 4.58 & 4.62 & 4.54 & 3.59 & 4.89 \\
R$_\mathrm{Magnetic}$ [\%] & 3.85 & 2.83 & 4.65 & 5.81 & 3.15
\end{tabular}
\end{ruledtabular}
\end{table*}

Our magnetic structure solution for the AFM I phase is different to that previously reported \cite{Prodi09}. In particular, we find both $A$ and $B$ sublattices ordering together below $T_1$, and both with a \textbf{k} = (0,0,0) propagation vector. Quantitative evidence for these differences may be found in our neutron diffraction data. The inset of Figure \ref{FIG:La_NPD_fig}b shows the $\{121\}$, $\{\bar{2}11\}$, and $\{\bar{1}12\}$ diffraction intensities in detail, which are mostly magnetic in orgin (paramagnetic data is shown in green). The magnetic structure factor of the $\{\bar{2}11\}$ and $\{\bar{1}12\}$ reflections is exactly zero for the collinear $B$-site order described above, and also zero for the similar $B$-site order reported in reference \citenum{Prodi09}. Furthermore, it is zero for each Mn4 and Mn5 sublattice, meaning that no degree of in-plane ferromagnetic canting between Mn4 and Mn5 sites, nor different magnitude moments of the two (both giving a net magnetic moment), can account for the observed intensity at these reflections. Out-of-plane antiferromagnetic canting of $B$ site moments (zero net moment) is allowed by symmetry, as mentioned above. In this case, finite magnetic diffraction intensities would be expected at the $\{\bar{2}11\}$ and $\{\bar{1}12\}$ reflections, but at the expense of the $\{121\}$ intensity whose structure factor is exactly zero for this additional $b$ axis component. Indeed, when freely refined, this out-of-plane component was found to be zero within error as it is inconsistent with other magnetic intensities found throughout the diffraction pattern. No other modifications of the collinear $B$-site structure are consistent with $\Gamma_2^+$ symmetry. However, the manganese $A$ site magnetic structure described above does have a non-zero structure factor for the $\{\bar{2}11\}$ and $\{\bar{1}12\}$ reflections, and, combined with the $B$ site order, faithfully reproduces all other diffraction intensities. Hence, given the basic $\chi$-type $B$ site magnetic structure, concomitant \textbf{k} = (0,0,0) $A$ and $B$ sublattice ordering is unambiguously demonstrated by our neutron diffraction experiments. Furthermore, the $A$ site structure naturally gives rise to an uncompensated 4$\mu_\mathrm{B}$ per formula unit, as observed in bulk magnetometry measurements (see Figure \ref{FIG:La_MvH}, below). 

Below $T_2$, three substantial changes in the neutron powder diffraction data occured simultaneously. Firstly, the relative intensities of the \textbf{k} = (0,0,0) magnetic diffraction peaks were modified, which could be modelled by a global tilt of the AFM I collinear magnetic structure towards the $b$ axis (Table \ref{TAB:Magnetic2}). The symmetry of this tilted structure is described by an admixture of $\Gamma_2^+$ and $\Gamma_1^+$ irreducible representations, which is allowed below the second magnetic phase transition. The tilt angle is plotted as a function of temperature in Figure \ref{FIG:La_temp_dep_fig}c, which indeed demonstrates critical behaviour at $T_2$. Secondly, an additional family of relatively strong, magnetic diffraction peaks appeared (labelled $\alpha$ in Figure \ref{FIG:La_NPD_fig}c), which could be unambiguously indexed with the propagation vector \textbf{k} = (0,1,0), as reported\cite{Prodi09}. 
By considering the full set of \textbf{k} = (0,1,0) reflections, systematic extinctions were observed at, for example, the $\{100\}$, $\{001\}$, $\{120\}$, and $\{021\}$ peak positions. Generic structure factor calculations demonstrated that these extinctions can only arise if the \textbf{k} = (0,1,0) component of the magnetic structure resides solely on the $B$ sublattice. The same calculations also demonstrated that these new magentic diffraction peaks are consistent with the $\alpha$ mode illustrated in Figure \ref{FIG:La_mag_struc_fig}b. Thirdly, two weak reflections were observed as highlighted in the inset of Figure \ref{FIG:La_NPD_fig}c (labelled $\gamma$), which also indexed with \textbf{k} = (0,1,0). These $\{$1 $\pm$1 1$\}$ and $\{\bar{1}$ $\pm$1 1$\}$ reflections are characteristic of a small $\gamma$ component of the $B$ sublattice magnetic structure (Figure \ref{FIG:La_mag_struc_fig}c).

\begin{table}
\caption{\label{TAB:Magnetic2} Magnetic structure parameters refined in the AFM II and AFM II' phases of \lmo\ and \nmo, respectively. Moment directions are given in spherical coordinates defined such that $m_z = m\cos(\theta)~||~c$, $m_x = m\cos(\phi)\sin(\theta)~||~a^*$, and $m_y = m\sin(\phi)\sin(\theta)~||~b$. Parameters given without standard errors have been fixed to maintain a magnetic structure with full moments on all sites. Mn1 and Mn3 moments are parallel, and antiparallel to Mn2 moments. Mn4 and Mn5 moments are given for sites $[\tfrac{1}{4},\tfrac{1}{4},\tfrac{1}{4}]$ and $[\tfrac{1}{4},\tfrac{1}{4},\tfrac{3}{4}]$, respectively. Moments on the other $B$ sites are transformed by the respective propagation vector.}
\begin{ruledtabular}
\begin{tabular}{l | c c c c }
 & \multicolumn{2}{c}{\lmo} & \multicolumn{2}{c}{\nmo}\\
\hline
$T$ [K] & \multicolumn{2}{c}{1.5} & \multicolumn{2}{c}{1.5} \\
\\
$R$ \hspace*{\fill} $m [\mu_\mathrm{B}]$ & \multicolumn{2}{c}{-} & \multicolumn{2}{c}{0.84(3)} \\
    \hspace*{\fill} $\phi [^\circ]$ & \multicolumn{2}{c}{-} & \multicolumn{2}{c}{314.6(5)} \\
    \hspace*{\fill} $\theta [^\circ]$ & \multicolumn{2}{c}{-} & \multicolumn{2}{c}{150.5(7)} \\
    \\
Mn1 \hspace*{\fill} $m [\mu_\mathrm{B}]$ & \multicolumn{2}{c}{3.86(2)} & \multicolumn{2}{c}{3.73(2)} \\
    \hspace*{\fill} $\phi [^\circ]$ & \multicolumn{2}{c}{323.7(3)} & \multicolumn{2}{c}{314.6(5)} \\
    \hspace*{\fill} $\theta [^\circ]$ & \multicolumn{2}{c}{-45.2(6)} & \multicolumn{2}{c}{-29.5(7)} \\
    \\
& $\chi$ & $\alpha + \gamma$ & $\chi$ & $\alpha + \gamma$ \\
Mn4 \hspace*{\fill} $m [\mu_\mathrm{B}]$ & 3.24(3) & 1.56(2) & 3.24(3) & 0.97(5) \\
    \hspace*{\fill} $\phi [^\circ]$ & 323.7(3) & -75.8 & 314.6(5) & -145(2) \\
    \hspace*{\fill} $\theta [^\circ]$ & -45.2(6) & 52.4(4) & -29.5(7) & 96.3 \\
    \\
Mn5 \hspace*{\fill} $m [\mu_\mathrm{B}]$ & 2.85 & 2.17(2) & 3.02 & 1.83(5) \\
    \hspace*{\fill} $\phi [^\circ]$ & 323.7(3) & -75.8 & 314.6(5) & -145(2) \\
    \hspace*{\fill} $\theta [^\circ]$ & 134.8(6) & 52.4(4) & 150.5(7) & 96.3 \\
    \\
\multicolumn{3}{l}{Fit reliability parameters}\\
R [\%] & \multicolumn{2}{c}{3.95} & \multicolumn{2}{c}{5.26} \\
$w$R [\%] & \multicolumn{2}{c}{4.48} & \multicolumn{2}{c}{5.16} \\
R$_\mathrm{Magnetic}$ [\%] & \multicolumn{2}{c}{2.65} & \multicolumn{2}{c}{3.87}
\end{tabular}
\end{ruledtabular}
\end{table}

Under the constraint that every magnetic ion should have a saturated magnetic moment in the ground state, the $\chi$ and ($\alpha+\gamma$) components must be orthogonal, giving rise to a noncollinear $B$ sublattice with an average spin direction collinear with the $A$ sublattice moments. In this case the small $\gamma$ component naturally arises if the noncollinearity is different for Mn4 and Mn5, which is indeed allowed by symmetry. This AFM II magnetic structure is shown in Figure \ref{FIG:La_mag_struc_fig}e, the refined parameters are given in Table \ref{TAB:Magnetic2}, and the temperature dependence of the $\alpha$ canting angle is shown in Figure \ref{FIG:La_temp_dep_fig}, which again displays critical behaviour at $T_2$. We note that in order to describe the symmetry of the AFM II magnetic structure it is necessary to mix $\mathrm{Y}_1^-$ and $\mathrm{Y}_2^-$ (see Table \ref{TAB:irreps} of the Appendix), as well as $\Gamma_1^+$ and $\Gamma_2^+$ irreducible representations. This combination corresponds to a substantial lowering of symmetry to a centrosymmetric triclinic magnetic space group.

\subsection{\label{NdSec}NdMn$_7$O$_{12}$}

\begin{figure}
\includegraphics[width=8.8cm]{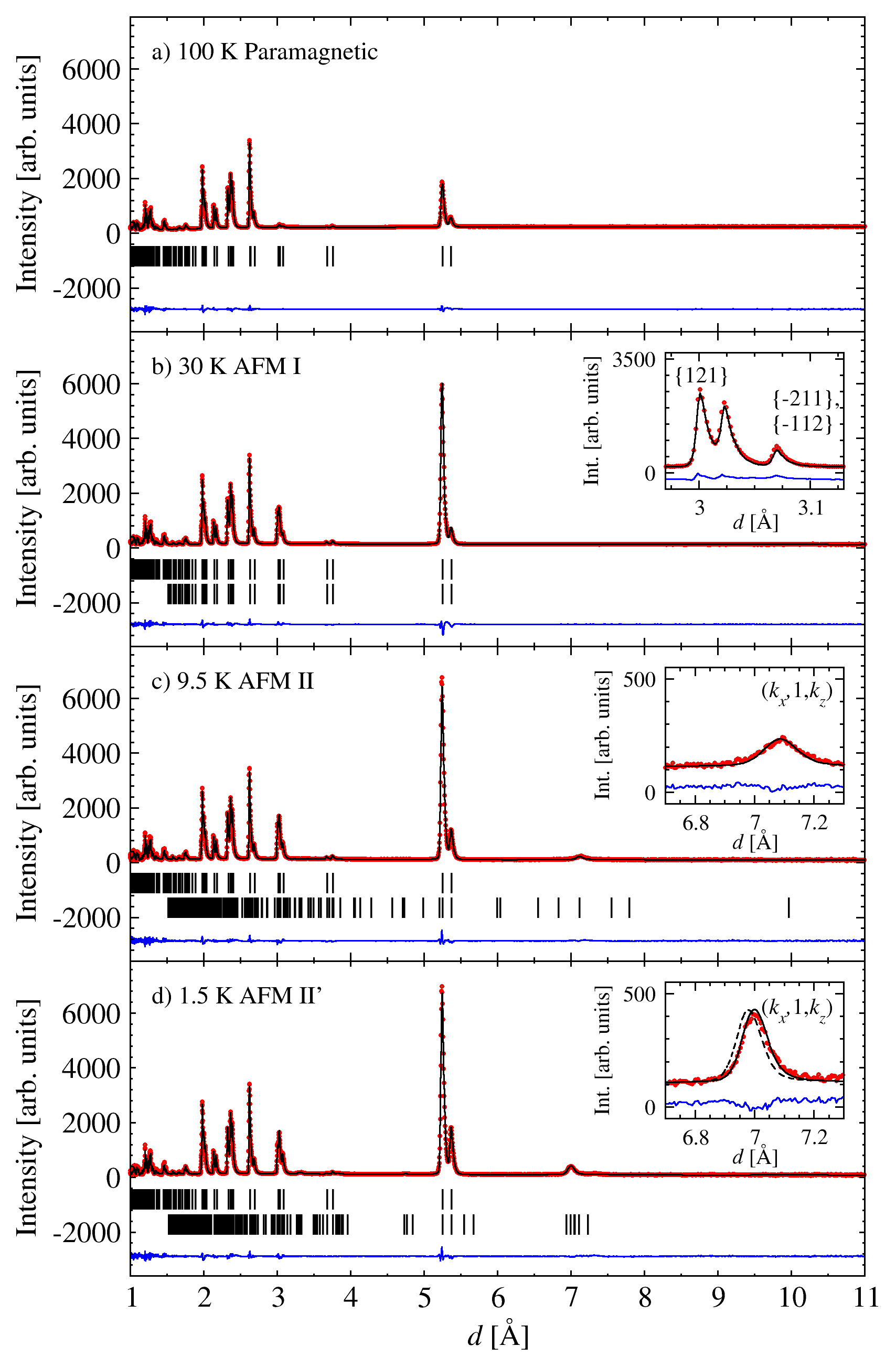}
\caption{\label{FIG:Nd_NPD_fig}Neutron powder diffraction data measured from \nmo\ in bank 2 (average $2\theta = 58.3^\circ$) of the WISH diffractometer. Diffraction data (red points) collected in the paramagnetic, AFM I, AFM II, and AFM II' phases are shown in panes a), b), c), and d), respectively. The fitted nuclear (top tick marks) and magnetic (bottom tick marks) structural models are shown as a solid black line. A difference pattern ($I_\mathrm{obs}-I_\mathrm{calc}$) is given as a blue solid line at the bottom of each pane. The dashed line in the inset to pane d shows the calculated peak position for the commensurate propagation vector (1/3,1,0).}
\end{figure}

\begin{figure}
\includegraphics[width=8.8cm]{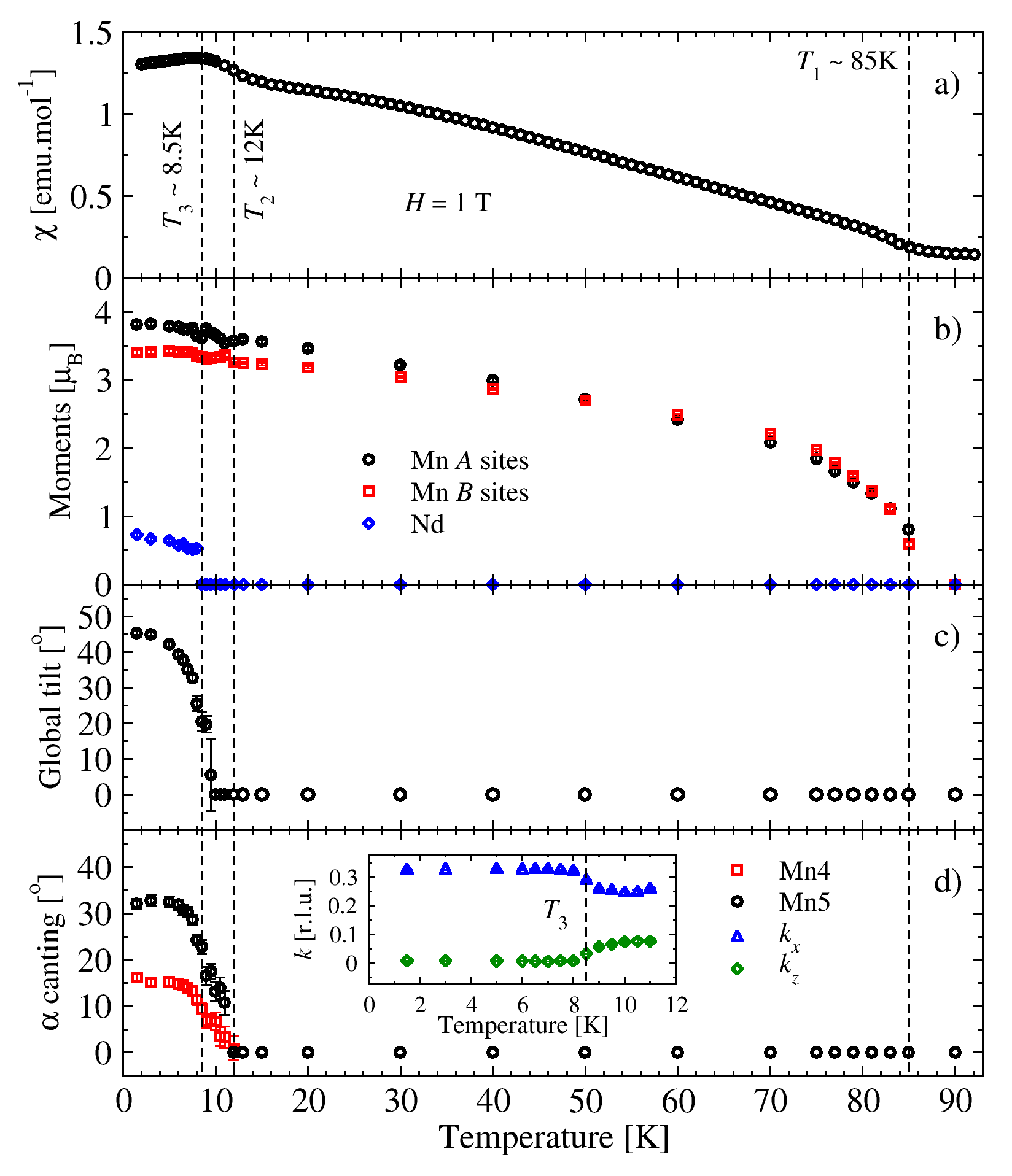}
\caption{\label{FIG:Nd_temp_dep_fig}The \nmo\ temperature dependence of a) field cooled magnetic susceptibility, b) the $A$ site, $B$ site, and Nd ordered magnetic moments, c) the global tilt of the \textbf{k}=(0,0,0) magnetic structure component, and d) the maximum canting angle of the Mn4 and Mn5 moments due to the evolution of the \textbf{k}=($k_x$,1,$k_z$) $\alpha$ mode, defined as $\tan^{-1}(m_\alpha/m_\chi)$. The inset of (d) gives the temperature dependence of the incommensurate propagation vector components, $k_x$ and $k_z$.}
\end{figure}

The field cooled magnetic susceptibility of \nmo, shown in Figure \ref{FIG:Nd_temp_dep_fig}a, demonstrated three magnetic phase transitions at $T_1 = 85$ K, $T_2 = 12$ K, and $T_3 = 8.5$ K. Neutron powder diffraction data collected at 100 K, in the paramagnetic phase of \nmo, (Figure \ref{FIG:Nd_NPD_fig}a) was fit with the same $I2/m$ crystal structure model as employed for \lmo. The model was found to be in excellent agreement with the data, and the refined structural parameters are given in Table \ref{TAB:Crystal}. Furthermore, this analysis demonstrated that our \nmo\ sample was 100\% phase pure within the sensitivity of the measurement.

Below $T_1$ additional magnetic diffraction intensities were observed, which could be indexed with the propagation vector \textbf{k}=(0,0,0); as was the case for \lmo. In fact, the diffraction patterns of both compounds below $T_1$ were found to be qualitatively identical (except for the small impurity peaks present for the La compound). The AFM I magnetic structure model was therefore fit to \nmo\ powder diffraction data measured at 30 K (Figure \ref{FIG:Nd_NPD_fig}b), and was found to accurately reproduce all magnetic intensities, including the $\{\bar{2}11\}$ and $\{\bar{1}12\}$ reflections that evidence $A$ site magnetic order (Figure \ref{FIG:Nd_NPD_fig}b inset). The refined AFM I magnetic structure parameters are given in Table \ref{TAB:Magnetic}.

\begin{figure}
\includegraphics[width=8.8cm]{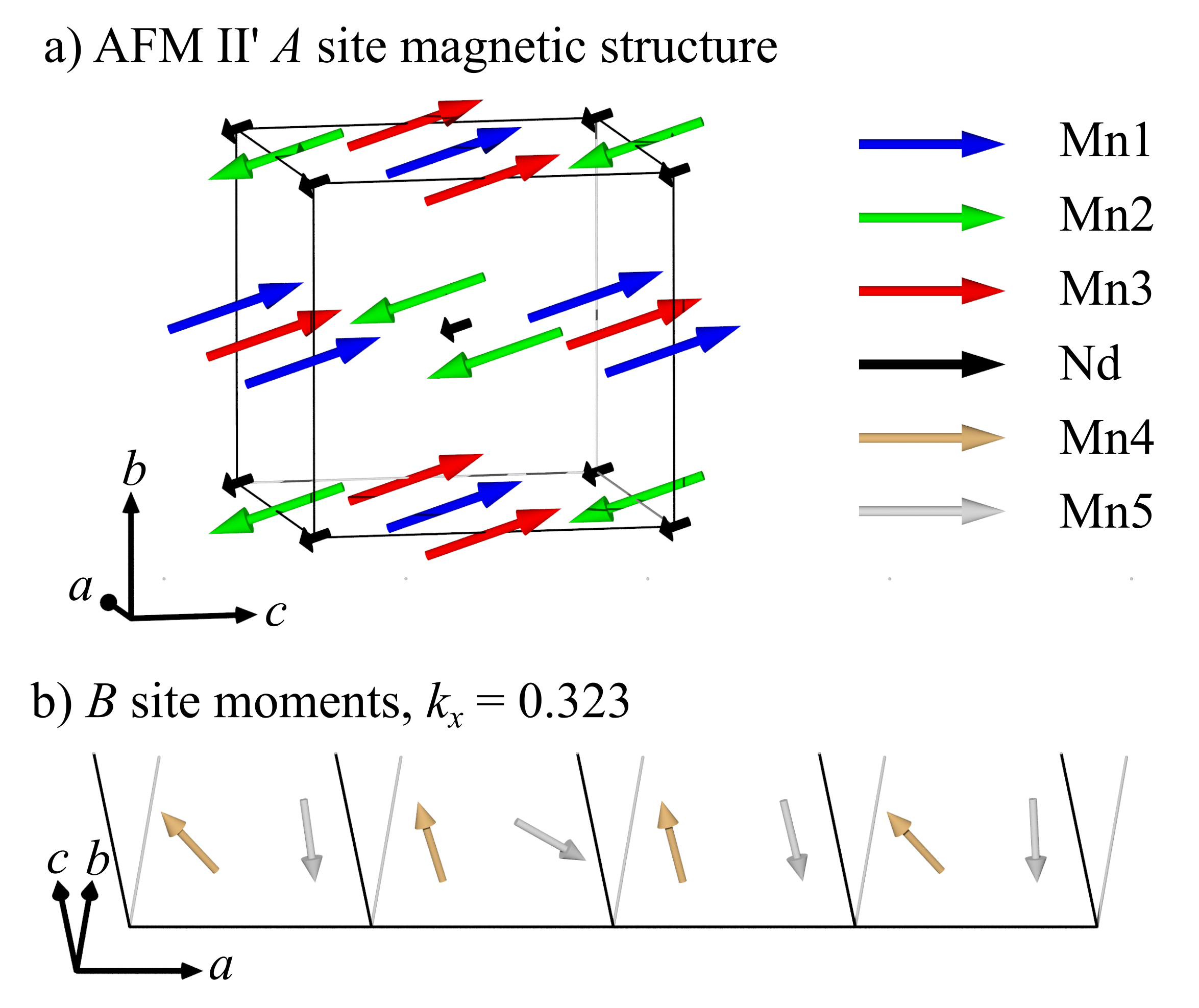}
\caption{\label{FIG:Nd_mag_struc_fig} a) The refined $A$ site magnetic structure of the AFM II' phase of \nmo. b) The incommensuarte `rocking' of $B$ site magnetic moments along the crystallographic $a$ axis ($\parallel k_x$). In all panes the $I2/m$ quadruple perovskite unit cell is drawn in faint grey lines.}
\end{figure}

In the temperature range $T_3<T<T_2$, additional magnetic diffraction peaks appeared in positions similar to those observed from \lmo\ below $T_2$. However, these peaks could not be indexed with the propagation vector \textbf{k}=(0,1,0), nor any other commensurate propagation vector corresponding to a high symmetry point of the $I2/m$ Brillouin zone. The highest symmetry incommensurate (ICM) propagation vectors of space group $I2/m$ are either parallel or perpendicular to $b^*$. Assuming that the ICM propagation vector was close to (0,1,0), the former scenario could be ruled out based upon the absence of (0,1$\pm k_y$,0) satellites. A systematic search of the reciprocal space plane ($k_x$,1,$k_z$) found that the propagation vector (0.248(2),1,0.064(3)) reproduced the positions of all magnetic diffraction intensities ($> 5$ observed). For completeness, the full Brillouin zone was searched and no other solutions were found. On cooling below $T_3$ there occured a significant shift in the positions of the ICM diffraction peaks, whilst their intensities continued to increase monotonically. At 1.5 K these peaks could be indexed with the propagation vector (0.3231(7),1,0.0069(7)) --- further supporting the ($k_x$,1,$k_z$) indexing solution. We note that this incommensurate propagation vector is close to the commensurate vector (1/3,1,0), however, the peak positions for \textbf{k}=(1/3,1,0) are inconsistent with the diffraction data (dashed line in Figure \ref{FIG:Nd_NPD_fig}d inset). Furthermore,  there is no symmetry reason for the propagation vector to lock in at (1/3,1,0) which is neither a high symmetry point in the Brillouin zone nor associated with any pseudo symmetry of the $I2/m$ crystal structure.

In both low temperature phases of \nmo\ the additional magnetic diffraction intensities were well fit by the same AFM II model as found for \lmo, but with the respective ICM propagation vectors assigned to the $(\alpha+\gamma)$ components residing on the $B$ sites (Figures \ref{FIG:Nd_NPD_fig}c and \ref{FIG:Nd_NPD_fig}d). To reinforce this similarity we label the $T_3<T<T_2$ phase of \nmo\ AFM II, and the ground state phase, AFM II'. As before, under the constraint that the magnetic structure has fully saturated moments, the ICM component must be oriented orthogonal to the \textbf{k} = (0,0,0) component. The $B$ site magnetic structure can then be understood either in terms of a long period planar `rocking' about the average \textbf{k} = (0,0,0) direction, or a conical `rotation' about the average. These two cases are described by an additional spin density wave or cycloidal modulation of the spins, respectively, which, as is typically the case, cannot be readily differentiated by neutron powder diffraction. The ground state magnetic structure of \nmo\ is illustrated in Figure \ref{FIG:Nd_mag_struc_fig} and parameterised in Table \ref{TAB:Magnetic2} assuming a rocking type structure. The temperature dependencies of the magnetic structure parameters are shown in Figure \ref{FIG:Nd_temp_dep_fig}, including the evolution of the ICM propagation vector.

To maintain a physical value of the Mn$^{3+}$ $A$ site magnetic moments below $T_3$, \emph{i.e.} $\leq 4 \mu_\mathrm{B}$, it was necessary to include \textbf{k}=(0,0,0) magnetic order on the Nd ions, as shown in Figure \ref{FIG:Nd_temp_dep_fig}b. Here, ferromagnetically aligned Nd moments of magnitude $0.84(3) \mu_\mathrm{B}$ were found to lie antiparallel to the uncompensated moment of the manganese $A$ site sublattice. It is well known that in rare earth manganites low temperature magnetic ordering of the rare earth sublattice can modify the magnetic order already present on the manganese sublattice. For example, in NdMnO$_3$ the Nd ions ferromagnetically order below 15 K with moments of $\sim 1 \mu_\mathrm{B}$, which drives a spin reorientation of the manganese sublattice and a net magnetisation reversal \cite{Kumar17}. In \nmo, the low temperature long range order of the Nd sublattice has two consequences. Firstly, it reduces the macroscopic ferrimagnetic moment, as evidenced in the field cooled magnetic susceptibility of \nmo\ (Figure \ref{FIG:Nd_temp_dep_fig}a). Secondly, Nd ordering can modify the effective exchange coupling between the $A$ and $B$ sublattices via additional $f$-$d$ exchange interactions. If $A$-$B$ coupling is pivotal in determining the ICM propagation vector, then the long range ordering of the Nd ions can naturally account for the changes in the manganese magnetic structure at $T_3$, which do not occur in the case of non-magnetic lanthanum.

\subsection{\label{CSESec}CeMn$_7$O$_{12}$, SmMn$_7$O$_{12}$, and EuMn$_7$O$_{12}$}

\begin{figure}
\includegraphics[width=8.5cm]{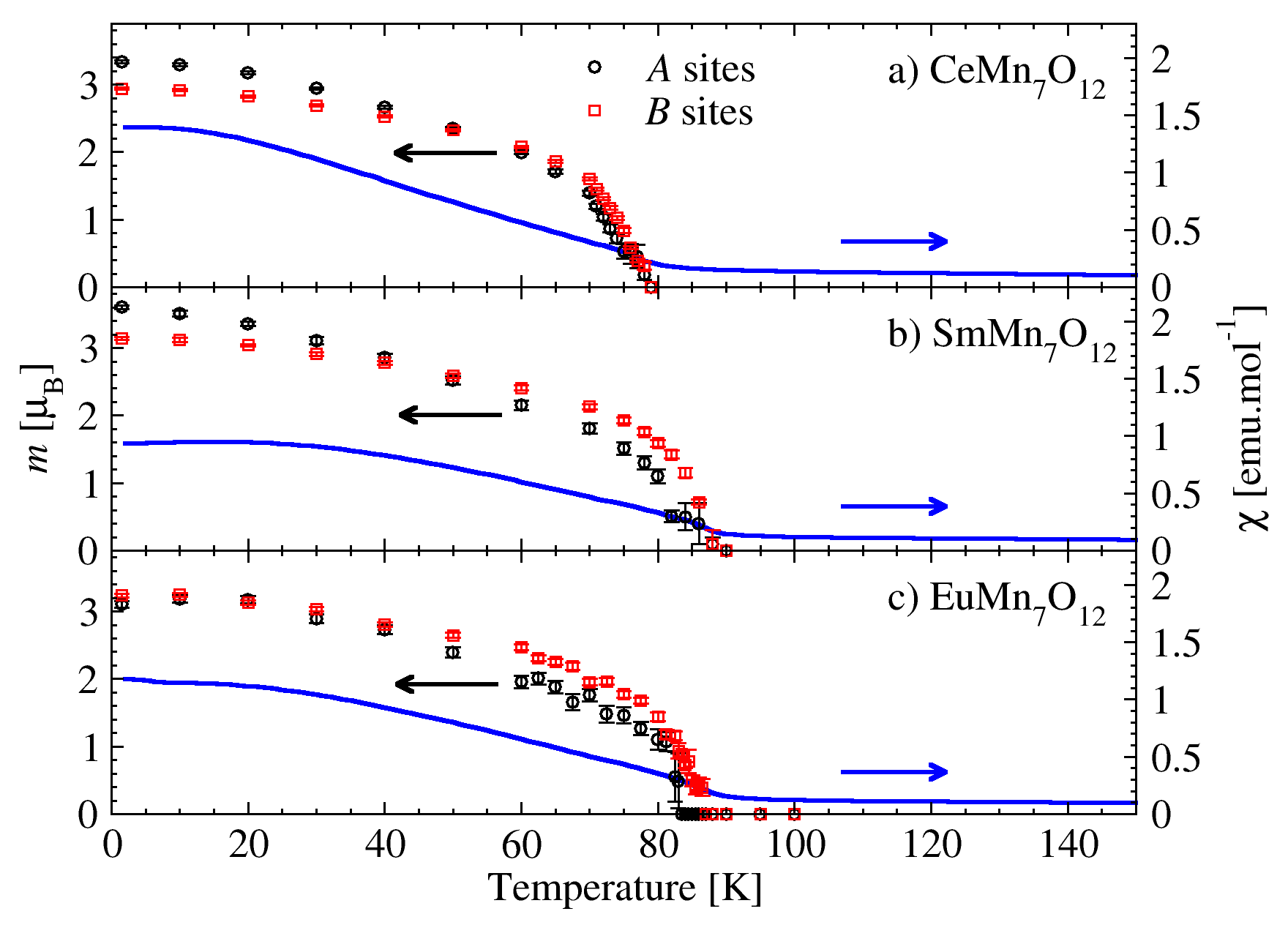}
\caption{\label{FIG:CSE_tempdep}The temperature dependence of the 1 T field cooled magnetic susceptibility (blue lines) and the AFM I $A$ and $B$ site magnetic moments (black cirlces and red squares, respectively) of \cmo, \smo, and \emo.}
\end{figure}

\begin{figure*}
\includegraphics[width=15cm]{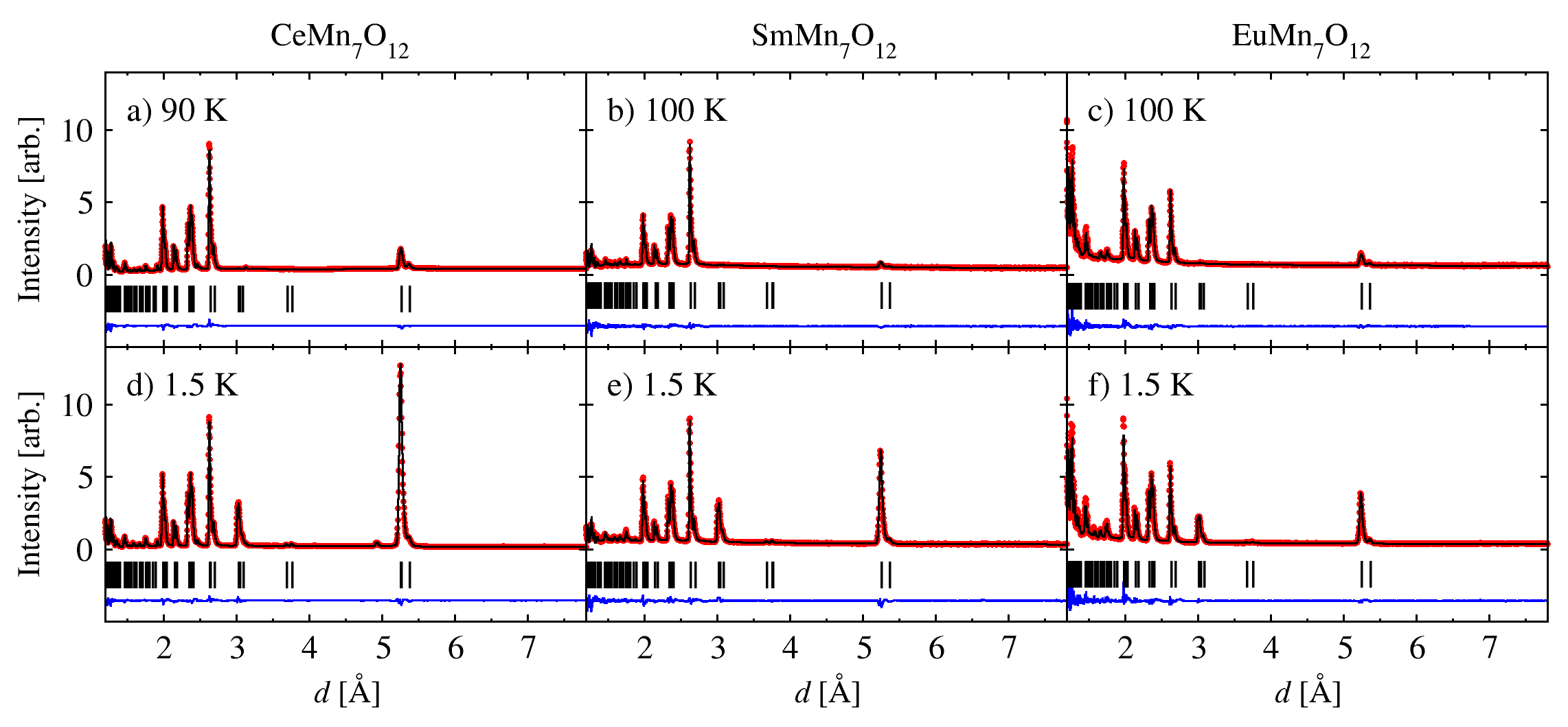}
\caption{\label{FIG:Combined_NPD_fig}Neutron powder diffraction data measured from \cmo, \smo, and \emo, in bank 2 (average $2\theta = 58.3^\circ$) of the WISH diffractometer. Diffraction data (red points) collected in the paramagnetic and AFM I phases are shown in panes (a-c) and (d-f), respectively. The fitted nuclear and magnetic structural models are shown as solid black lines, with the respective peak positions shown by a single sets of black tick marks. A difference pattern ($I_\mathrm{obs}-I_\mathrm{calc}$) is given as a blue solid line at the bottom of each pane. Tick marks from impurity phases have been omitted for clarity.}
\end{figure*}

Magnetic susceptibility measurements of compounds with $R$ = Ce, Sm, and Eu, shown in Figure \ref{FIG:CSE_tempdep}, demonstrated a single magnetic phase transition; for \cmo\ $T_1$ = 80 K, and for \smo\ and \emo\, $T_1$ = 87 K. Neutron powder diffraction data collected in the paramagnetic phase of all three compounds (Figure \ref{FIG:Combined_NPD_fig} a-c) was reliably fit with the same $I2/m$ crystal structure model as employed for the $R$ = La and Nd compounds, and the structural parameters are given in Table \ref{TAB:Crystal}. The \smo\ and \emo\ samples were found to be 100\% phase pure. \cmo\ was found to be 96\% pure, with 3\% CeO$_2$ and 1\% Mn$_3$O$_4$ impurities present. Furthermore, the \cmo\ phase was found to be non-stoichiometric, with a refined composition of Ce$_{0.86}$Mn$_{7.14}$O$_{12}$. In this case Mn$^{2+}$ ions were heterovalently substituted for Ce$^{3+}$ ions, as found for (Tb$_{0.88}$Mn$_{0.12}$)Mn$_7$O$_{12}$ \cite{Zhang18}, which leads to hole doping on the $B$ sites.

For all three compounds ($R$ = Ce, Sm, and Eu), variable temperature neutron powder diffraction experiments confirmed a single magnetic phase from $T_1$ down to 1.5 K. Magnetic diffraction intensities appeared below $T_1$, which could be fit with the \textbf{k}=(0,0,0) AFM I model (Figure \ref{FIG:Combined_NPD_fig} d-f). The respective magnetic structure parameters refined at 1.5 K are given in Table \ref{TAB:Magnetic}, and the temperature dependencies of the $A$ and $B$ site moments are shown in Figure \ref{FIG:CSE_tempdep}. For the \smo\ and \emo\ data analysis an absorption correction was required, based upon a cylindrical geometry as implemented in Fullprof \cite{Rodriguezcarvaja93}

\section{\label{dissec}Discussion}

As described theoretically by Goodenough, Kanamori, and Anderson\cite{anderson50,goodenough55,kanamori57,goodenough58}, orbital order is widely considered to be the primary factor in mediating magnetic exchange in manganese based perovskite oxides and their variants \cite{Wollan1955,Radaelli97,tokura06,Johnson17}. In $R$Mn$_7$O$_{12}$, the orthogonal in-plane ordering of $B$ site Mn$^{3+}$ $d_{3x^2-r^2}$ and $d_{3z^2-r^2}$ orbitals favours ferromagnetic exchange within the $ac$ planes, which are coupled antiferromagnetically along $y$ ($\alpha$ mode), as observed in LaMnO$_3$ \cite{Wollan1955}. It is surprising, therefore, to find that all measured $R$Mn$_7$O$_{12}$ quadruple perovksites order below $T_1$ with a $B$ site magnetic structure of antiferromagnetically coupled moments in the $ac$ plane, which are ferromagnetically coupled along $b$ ($\chi$ mode) --- the exact opposite. Based on the bonding geometry of the perovskite crystal structure one might expect $B$-$B$ interactions to dominate the magnetic exchange energy. However, in $R$Mn$_7$O$_{12}$ the bonding geometry is highly distorted owing to the octahedral tilt pattern that accommodates the ordered occupation of the $A$ site cations. As a result, the $B$-O-$B$ bonds may be close the critical angle between ferromagnetic and antiferromagnetic exchange, giving rise to the possibility that they are in fact weak with respect to $A$-$A$ and $A$-$B$ exchange interactions. The latter interactions would then play the lead role in determining the experimentally observed magnetic structure at the expense of the $B$-$B$ exchange mediated by orbital order. In \lmo\ and \nmo\ an instability towards $\alpha$ type $B$ site magnetic order is apparent below the low temperature phase transitions, which indicates that magnetic frustration due to orbital order plays a key role in the stability of the ground state magnetic structures.

\begin{figure}
\includegraphics[width=8.5cm]{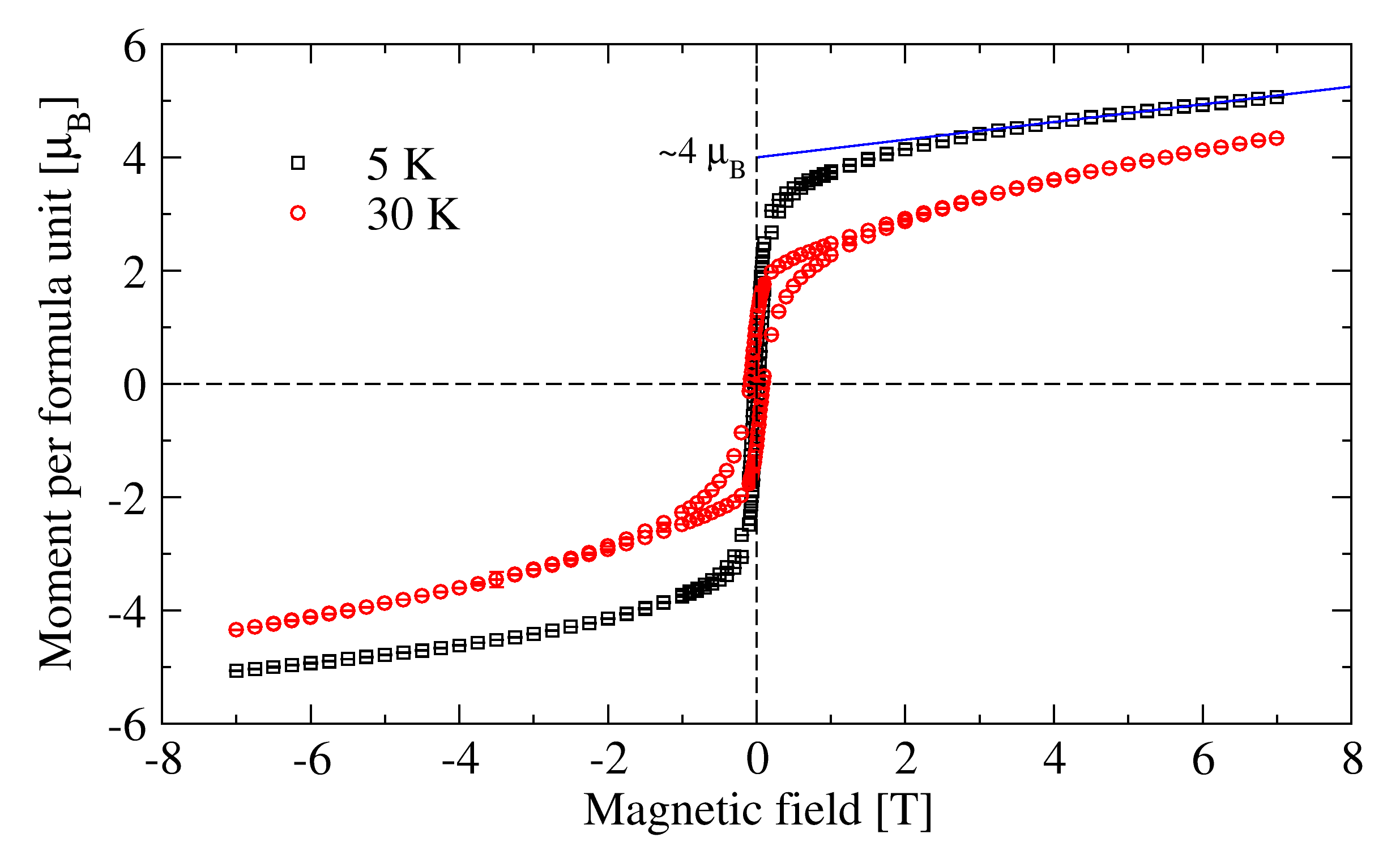}
\caption{\label{FIG:La_MvH}The magnetic moment per formula unit of our polycrystalline \lmo\ sample, measured as a function of magnetic field above and below $T_2$.}
\end{figure}

Our neutron powder diffraction results on the evolution of the $R$Mn$_7$O$_{12}$ magnetic structures are further supported by bulk magnetisation measurements. Figure \ref{FIG:La_MvH} shows the magnetic moment per formula unit of a pressed pellet of our polycrystalline \lmo\ sample, measured as a function of applied magnetic field. Above $T_2$, a remnant moment of approximately 2 $\mu_\mathrm{B}$ per formula unit was observed in the powder averaged data. Below $T_2$, the hysteresis loop narrowed, and the remnant moment increased to 4 $\mu_\mathrm{B}$ per formula unit --- the full uncompensated moment of the $A$ site ferrimagnetic sublattice. This behaviour is consistent with the temperature dependent susceptibility shown in Figure \ref{FIG:La_temp_dep_fig}a. In pseudocubic symmetry the collinear magnetic structure of the $A$ sublattice is isotropic. In the monoclinic phase this sublattice becomes only weakly anisotropic owing to relatively small distortions away from cubic symmetry. Hence, the magnetic anisotropy of the uncompensated $A$ sublattice moment is likely inherited from coupling to the strongly anisotropic $B$-sites. This $A$-$B$ coupling exists only for the \textbf{k}=(0,0,0) components, by symmetry. The inhereted $A$ sublattice anisotropy will therefore be reduced upon $\alpha$ canting of the $B$ sites below $T_2$, giving rise to the apparent increase in remnant magnetisation in the powder averaged data of Figure \ref{FIG:La_MvH}. Regarding magnetic anisotropy, one important question remains unanswered: Which anisotropy determines the moment directions in the ground state of \lmo\ and \nmo, giving rise to the global tilt of the magnetic structures?

Finally, it is noteworthy that the low temperature phase transitions to non-collinear ground states have only been observed in compounds with large rare earth ionic radius (La and Nd). However, \cmo\ appears to be an exception; the ionic radius of Ce is intermediate with respect to La and Nd, yet no low temperature transition was observed. Our \cmo\ sample was found to be non-stoichiometric, giving rise to hole doping on the Mn $B$ sites. Such doping will diminish the orbital order of the $B$ sites, which is consistent with the suppression of a non-collinear ground state that is otherwise established by magnetic frustration induced by orbital order. To test this observation we measured a non-stoichiometric sample of \lmo\ (composition La$_{0.9}$Mn$_{7.1}$O$_{12}$), in which we observed weak anomalies in the heat capacity and magnetic susceptibility at $T_2$, and broad \textbf{k}=(0,1,0) magnetic diffraction peaks below $T_2$ (not shown here). Both observations are evidence of the suppression of the non-collinear ground state. Furthermore, variable temperature neutron powder diffraction data measured from La$_{0.9}$Mn$_{7.1}$O$_{12}$, also not shown here, demonstrated a reduction of $T_1$ by 3.5 K, which was also observed upon increasing $x$ in (Tb$_{1-x}$Mn$_x$)Mn$_7$O$_{12}$\cite{Zhang18}.

\section{\label{consec}Conclusions}

In all measured $R$Mn$_7$O$_{12}$ compounds we have found that the $A$ and $B$ manganese sublattices magnetically order concomitantly on coooling below the first magnetic phase transition at $T_1$ (between 80 and 90 K). The magnetic structure (labelled AFM I throughout) is collinear, with one uncompensated $A$ site Mn$^{3+}$ magnetic moment per formula unit. This uncompensated moment naturally explains the bulk magnetisation measured throughout the magnetic phases. In \lmo\ and \nmo\ a second phase transition occurs at low temperature ($T_2$), which we have shown to be related to the onset of a secondary component of the magentic structure associated with the $B$ site sublattice. In \lmo\ this component is commensurate, and in \nmo\ it is incommensurate, but in both cases it likely reflects a magnetic instability originating in the $B$ site orbital order. In this regard the $R$Mn$_7$O$_{12}$ compounds appear to be somewhat unconventional, as the magnetic order is not primarily determined by the underlying orbital order, as would be expected in the manganites. Our results might be explored further in both theoretical and experimental studies of the magnetic exchange interactions of $R$Mn$_7$O$_{12}$, and the microscopic distinctions between these materials and, for example, the A$^{1+,2+}$Mn$_7$O$_{12}$ quadruple perovskites and the $R$MnO$_3$ simple perovskites.

\begin{acknowledgments}
RDJ acknowledges support from a Royal Society University Research Fellowship, and fruitful discussions with Dr. F. Orlandi. The work performed in Japan was supported in part by JSPS KAKENHI, Grant Numbers JP15K14133 and JP16H04501.
\end{acknowledgments}

\bibliography{rmo}

\begin{thebibliography}{33}%
\makeatletter
\providecommand \@ifxundefined [1]{%
 \@ifx{#1\undefined}
}%
\providecommand \@ifnum [1]{%
 \ifnum #1\expandafter \@firstoftwo
 \else \expandafter \@secondoftwo
 \fi
}%
\providecommand \@ifx [1]{%
 \ifx #1\expandafter \@firstoftwo
 \else \expandafter \@secondoftwo
 \fi
}%
\providecommand \natexlab [1]{#1}%
\providecommand \enquote  [1]{``#1''}%
\providecommand \bibnamefont  [1]{#1}%
\providecommand \bibfnamefont [1]{#1}%
\providecommand \citenamefont [1]{#1}%
\providecommand \href@noop [0]{\@secondoftwo}%
\providecommand \href [0]{\begingroup \@sanitize@url \@href}%
\providecommand \@href[1]{\@@startlink{#1}\@@href}%
\providecommand \@@href[1]{\endgroup#1\@@endlink}%
\providecommand \@sanitize@url [0]{\catcode `\\12\catcode `\$12\catcode
  `\&12\catcode `\#12\catcode `\^12\catcode `\_12\catcode `\%12\relax}%
\providecommand \@@startlink[1]{}%
\providecommand \@@endlink[0]{}%
\providecommand \url  [0]{\begingroup\@sanitize@url \@url }%
\providecommand \@url [1]{\endgroup\@href {#1}{\urlprefix }}%
\providecommand \urlprefix  [0]{URL }%
\providecommand \Eprint [0]{\href }%
\providecommand \doibase [0]{http://dx.doi.org/}%
\providecommand \selectlanguage [0]{\@gobble}%
\providecommand \bibinfo  [0]{\@secondoftwo}%
\providecommand \bibfield  [0]{\@secondoftwo}%
\providecommand \translation [1]{[#1]}%
\providecommand \BibitemOpen [0]{}%
\providecommand \bibitemStop [0]{}%
\providecommand \bibitemNoStop [0]{.\EOS\space}%
\providecommand \EOS [0]{\spacefactor3000\relax}%
\providecommand \BibitemShut  [1]{\csname bibitem#1\endcsname}%
\let\auto@bib@innerbib\@empty
\bibitem [{\citenamefont {Vasil’ev}\ and\ \citenamefont
  {Volkova}(2007)}]{Vasiliev07}%
  \BibitemOpen
  \bibfield  {author} {\bibinfo {author} {\bibfnamefont {A.~N.}\ \bibnamefont
  {Vasil’ev}}\ and\ \bibinfo {author} {\bibfnamefont {O.~S.}\ \bibnamefont
  {Volkova}},\ }\href {\doibase 10.1063/1.2747047} {\bibfield  {journal}
  {\bibinfo  {journal} {Low Temperature Physics}\ }\textbf {\bibinfo {volume}
  {33}},\ \bibinfo {pages} {895} (\bibinfo {year} {2007})}\BibitemShut
  {NoStop}%
\bibitem [{\citenamefont {Gilioli}\ and\ \citenamefont
  {Ehm}(2014)}]{Gilioli14}%
  \BibitemOpen
  \bibfield  {author} {\bibinfo {author} {\bibfnamefont {E.}~\bibnamefont
  {Gilioli}}\ and\ \bibinfo {author} {\bibfnamefont {L.}~\bibnamefont {Ehm}},\
  }\href {\doibase 10.1107/S2052252514020569} {\bibfield  {journal} {\bibinfo
  {journal} {IUCrJ}\ }\textbf {\bibinfo {volume} {1}},\ \bibinfo {pages} {590}
  (\bibinfo {year} {2014})}\BibitemShut {NoStop}%
\bibitem [{\citenamefont {Zeng}\ \emph {et~al.}(1999)\citenamefont {Zeng},
  \citenamefont {Greenblatt}, \citenamefont {Subramanian},\ and\ \citenamefont
  {Croft}}]{Zeng99}%
  \BibitemOpen
  \bibfield  {author} {\bibinfo {author} {\bibfnamefont {Z.}~\bibnamefont
  {Zeng}}, \bibinfo {author} {\bibfnamefont {M.}~\bibnamefont {Greenblatt}},
  \bibinfo {author} {\bibfnamefont {M.~A.}\ \bibnamefont {Subramanian}}, \ and\
  \bibinfo {author} {\bibfnamefont {M.}~\bibnamefont {Croft}},\ }\href
  {\doibase 10.1103/PhysRevLett.82.3164} {\bibfield  {journal} {\bibinfo
  {journal} {Phys. Rev. Lett.}\ }\textbf {\bibinfo {volume} {82}},\ \bibinfo
  {pages} {3164} (\bibinfo {year} {1999})}\BibitemShut {NoStop}%
\bibitem [{\citenamefont {Mezzadri}\ \emph
  {et~al.}(2009{\natexlab{a}})\citenamefont {Mezzadri}, \citenamefont
  {Calestani}, \citenamefont {Calicchio}, \citenamefont {Gilioli},
  \citenamefont {Bolzoni}, \citenamefont {Cabassi}, \citenamefont {Marezio},\
  and\ \citenamefont {Migliori}}]{Mezzadri09_1}%
  \BibitemOpen
  \bibfield  {author} {\bibinfo {author} {\bibfnamefont {F.}~\bibnamefont
  {Mezzadri}}, \bibinfo {author} {\bibfnamefont {G.}~\bibnamefont {Calestani}},
  \bibinfo {author} {\bibfnamefont {M.}~\bibnamefont {Calicchio}}, \bibinfo
  {author} {\bibfnamefont {E.}~\bibnamefont {Gilioli}}, \bibinfo {author}
  {\bibfnamefont {F.}~\bibnamefont {Bolzoni}}, \bibinfo {author} {\bibfnamefont
  {R.}~\bibnamefont {Cabassi}}, \bibinfo {author} {\bibfnamefont
  {M.}~\bibnamefont {Marezio}}, \ and\ \bibinfo {author} {\bibfnamefont
  {A.}~\bibnamefont {Migliori}},\ }\href {\doibase 10.1103/PhysRevB.79.100106}
  {\bibfield  {journal} {\bibinfo  {journal} {Phys. Rev. B}\ }\textbf {\bibinfo
  {volume} {79}},\ \bibinfo {pages} {100106} (\bibinfo {year}
  {2009}{\natexlab{a}})}\BibitemShut {NoStop}%
\bibitem [{\citenamefont {Johnson}\ \emph {et~al.}(2012)\citenamefont
  {Johnson}, \citenamefont {Chapon}, \citenamefont {Khalyavin}, \citenamefont
  {Manuel}, \citenamefont {Radaelli},\ and\ \citenamefont
  {Martin}}]{Johnson12}%
  \BibitemOpen
  \bibfield  {author} {\bibinfo {author} {\bibfnamefont {R.~D.}\ \bibnamefont
  {Johnson}}, \bibinfo {author} {\bibfnamefont {L.~C.}\ \bibnamefont {Chapon}},
  \bibinfo {author} {\bibfnamefont {D.~D.}\ \bibnamefont {Khalyavin}}, \bibinfo
  {author} {\bibfnamefont {P.}~\bibnamefont {Manuel}}, \bibinfo {author}
  {\bibfnamefont {P.~G.}\ \bibnamefont {Radaelli}}, \ and\ \bibinfo {author}
  {\bibfnamefont {C.}~\bibnamefont {Martin}},\ }\href {\doibase
  10.1103/PhysRevLett.108.067201} {\bibfield  {journal} {\bibinfo  {journal}
  {Phys. Rev. Lett.}\ }\textbf {\bibinfo {volume} {108}},\ \bibinfo {pages}
  {067201} (\bibinfo {year} {2012})}\BibitemShut {NoStop}%
\bibitem [{\citenamefont {Marezio}\ \emph {et~al.}(1973)\citenamefont
  {Marezio}, \citenamefont {Dernier}, \citenamefont {Chenavas},\ and\
  \citenamefont {Joubert}}]{Marezio73}%
  \BibitemOpen
  \bibfield  {author} {\bibinfo {author} {\bibfnamefont {M.}~\bibnamefont
  {Marezio}}, \bibinfo {author} {\bibfnamefont {P.}~\bibnamefont {Dernier}},
  \bibinfo {author} {\bibfnamefont {J.}~\bibnamefont {Chenavas}}, \ and\
  \bibinfo {author} {\bibfnamefont {J.}~\bibnamefont {Joubert}},\ }\href
  {\doibase http://dx.doi.org/10.1016/0022-4596(73)90200-4} {\bibfield
  {journal} {\bibinfo  {journal} {Journal of Solid State Chemistry}\ }\textbf
  {\bibinfo {volume} {6}},\ \bibinfo {pages} {16 } (\bibinfo {year}
  {1973})}\BibitemShut {NoStop}%
\bibitem [{\citenamefont {Bochu}\ \emph {et~al.}(1974)\citenamefont {Bochu},
  \citenamefont {Chenavas}, \citenamefont {Joubert},\ and\ \citenamefont
  {Marezio}}]{BOCHU74}%
  \BibitemOpen
  \bibfield  {author} {\bibinfo {author} {\bibfnamefont {B.}~\bibnamefont
  {Bochu}}, \bibinfo {author} {\bibfnamefont {J.}~\bibnamefont {Chenavas}},
  \bibinfo {author} {\bibfnamefont {J.}~\bibnamefont {Joubert}}, \ and\
  \bibinfo {author} {\bibfnamefont {M.}~\bibnamefont {Marezio}},\ }\href
  {\doibase https://doi.org/10.1016/0022-4596(74)90102-9} {\bibfield  {journal}
  {\bibinfo  {journal} {Journal of Solid State Chemistry}\ }\textbf {\bibinfo
  {volume} {11}},\ \bibinfo {pages} {88 } (\bibinfo {year} {1974})}\BibitemShut
  {NoStop}%
\bibitem [{\citenamefont {Prodi}\ \emph {et~al.}(2009)\citenamefont {Prodi},
  \citenamefont {Gilioli}, \citenamefont {Cabassi}, \citenamefont {Bolzoni},
  \citenamefont {Licci}, \citenamefont {Huang}, \citenamefont {Lynn},
  \citenamefont {Affronte}, \citenamefont {Gauzzi},\ and\ \citenamefont
  {Marezio}}]{Prodi09}%
  \BibitemOpen
  \bibfield  {author} {\bibinfo {author} {\bibfnamefont {A.}~\bibnamefont
  {Prodi}}, \bibinfo {author} {\bibfnamefont {E.}~\bibnamefont {Gilioli}},
  \bibinfo {author} {\bibfnamefont {R.}~\bibnamefont {Cabassi}}, \bibinfo
  {author} {\bibfnamefont {F.}~\bibnamefont {Bolzoni}}, \bibinfo {author}
  {\bibfnamefont {F.}~\bibnamefont {Licci}}, \bibinfo {author} {\bibfnamefont
  {Q.}~\bibnamefont {Huang}}, \bibinfo {author} {\bibfnamefont {J.~W.}\
  \bibnamefont {Lynn}}, \bibinfo {author} {\bibfnamefont {M.}~\bibnamefont
  {Affronte}}, \bibinfo {author} {\bibfnamefont {A.}~\bibnamefont {Gauzzi}}, \
  and\ \bibinfo {author} {\bibfnamefont {M.}~\bibnamefont {Marezio}},\ }\href
  {\doibase 10.1103/PhysRevB.79.085105} {\bibfield  {journal} {\bibinfo
  {journal} {Phys. Rev. B}\ }\textbf {\bibinfo {volume} {79}},\ \bibinfo
  {pages} {085105} (\bibinfo {year} {2009})}\BibitemShut {NoStop}%
\bibitem [{\citenamefont {Mezzadri}\ \emph
  {et~al.}(2009{\natexlab{b}})\citenamefont {Mezzadri}, \citenamefont
  {Calicchio}, \citenamefont {Gilioli}, \citenamefont {Cabassi}, \citenamefont
  {Bolzoni}, \citenamefont {Calestani},\ and\ \citenamefont
  {Bissoli}}]{Mezzadri09_2}%
  \BibitemOpen
  \bibfield  {author} {\bibinfo {author} {\bibfnamefont {F.}~\bibnamefont
  {Mezzadri}}, \bibinfo {author} {\bibfnamefont {M.}~\bibnamefont {Calicchio}},
  \bibinfo {author} {\bibfnamefont {E.}~\bibnamefont {Gilioli}}, \bibinfo
  {author} {\bibfnamefont {R.}~\bibnamefont {Cabassi}}, \bibinfo {author}
  {\bibfnamefont {F.}~\bibnamefont {Bolzoni}}, \bibinfo {author} {\bibfnamefont
  {G.}~\bibnamefont {Calestani}}, \ and\ \bibinfo {author} {\bibfnamefont
  {F.}~\bibnamefont {Bissoli}},\ }\href {\doibase 10.1103/PhysRevB.79.014420}
  {\bibfield  {journal} {\bibinfo  {journal} {Phys. Rev. B}\ }\textbf {\bibinfo
  {volume} {79}},\ \bibinfo {pages} {014420} (\bibinfo {year}
  {2009}{\natexlab{b}})}\BibitemShut {NoStop}%
\bibitem [{\citenamefont {Locherer}\ \emph {et~al.}(2012)\citenamefont
  {Locherer}, \citenamefont {Dinnebier}, \citenamefont {Kremer}, \citenamefont
  {Greenblatt},\ and\ \citenamefont {Jansen}}]{Locherer12}%
  \BibitemOpen
  \bibfield  {author} {\bibinfo {author} {\bibfnamefont {T.}~\bibnamefont
  {Locherer}}, \bibinfo {author} {\bibfnamefont {R.}~\bibnamefont {Dinnebier}},
  \bibinfo {author} {\bibfnamefont {R.}~\bibnamefont {Kremer}}, \bibinfo
  {author} {\bibfnamefont {M.}~\bibnamefont {Greenblatt}}, \ and\ \bibinfo
  {author} {\bibfnamefont {M.}~\bibnamefont {Jansen}},\ }\href {\doibase
  http://dx.doi.org/10.1016/j.jssc.2012.02.036} {\bibfield  {journal} {\bibinfo
   {journal} {Journal of Solid State Chemistry}\ }\textbf {\bibinfo {volume}
  {190}},\ \bibinfo {pages} {277 } (\bibinfo {year} {2012})}\BibitemShut
  {NoStop}%
\bibitem [{\citenamefont {Ovsyannikov}\ \emph {et~al.}(2013)\citenamefont
  {Ovsyannikov}, \citenamefont {Abakumov}, \citenamefont {Tsirlin},
  \citenamefont {Schnelle}, \citenamefont {Egoavil}, \citenamefont {Verbeeck},
  \citenamefont {Van Tendeloo}, \citenamefont {Glazyrin}, \citenamefont
  {Hanfland},\ and\ \citenamefont {Dubrovinsky}}]{Ovsyannikov13}%
  \BibitemOpen
  \bibfield  {author} {\bibinfo {author} {\bibfnamefont {S.~V.}\ \bibnamefont
  {Ovsyannikov}}, \bibinfo {author} {\bibfnamefont {A.~M.}\ \bibnamefont
  {Abakumov}}, \bibinfo {author} {\bibfnamefont {A.~A.}\ \bibnamefont
  {Tsirlin}}, \bibinfo {author} {\bibfnamefont {W.}~\bibnamefont {Schnelle}},
  \bibinfo {author} {\bibfnamefont {R.}~\bibnamefont {Egoavil}}, \bibinfo
  {author} {\bibfnamefont {J.}~\bibnamefont {Verbeeck}}, \bibinfo {author}
  {\bibfnamefont {G.}~\bibnamefont {Van Tendeloo}}, \bibinfo {author}
  {\bibfnamefont {K.~V.}\ \bibnamefont {Glazyrin}}, \bibinfo {author}
  {\bibfnamefont {M.}~\bibnamefont {Hanfland}}, \ and\ \bibinfo {author}
  {\bibfnamefont {L.}~\bibnamefont {Dubrovinsky}},\ }\href {\doibase
  10.1002/anie.201208553} {\bibfield  {journal} {\bibinfo  {journal}
  {Angewandte Chemie International Edition}\ }\textbf {\bibinfo {volume}
  {52}},\ \bibinfo {pages} {1494} (\bibinfo {year} {2013})}\BibitemShut
  {NoStop}%
\bibitem [{\citenamefont {Glazkova}\ \emph {et~al.}(2015)\citenamefont
  {Glazkova}, \citenamefont {Terada}, \citenamefont {Matsushita}, \citenamefont
  {Katsuya}, \citenamefont {Tanaka}, \citenamefont {Sobolev}, \citenamefont
  {Presniakov},\ and\ \citenamefont {Belik}}]{Glazkova15}%
  \BibitemOpen
  \bibfield  {author} {\bibinfo {author} {\bibfnamefont {Y.~S.}\ \bibnamefont
  {Glazkova}}, \bibinfo {author} {\bibfnamefont {N.}~\bibnamefont {Terada}},
  \bibinfo {author} {\bibfnamefont {Y.}~\bibnamefont {Matsushita}}, \bibinfo
  {author} {\bibfnamefont {Y.}~\bibnamefont {Katsuya}}, \bibinfo {author}
  {\bibfnamefont {M.}~\bibnamefont {Tanaka}}, \bibinfo {author} {\bibfnamefont
  {A.~V.}\ \bibnamefont {Sobolev}}, \bibinfo {author} {\bibfnamefont {I.~A.}\
  \bibnamefont {Presniakov}}, \ and\ \bibinfo {author} {\bibfnamefont {A.~A.}\
  \bibnamefont {Belik}},\ }\href {\doibase 10.1021/acs.inorgchem.5b01472}
  {\bibfield  {journal} {\bibinfo  {journal} {Inorganic Chemistry}\ }\textbf
  {\bibinfo {volume} {54}},\ \bibinfo {pages} {9081} (\bibinfo {year}
  {2015})}\BibitemShut {NoStop}%
\bibitem [{\citenamefont {Belik}\ \emph {et~al.}(2016)\citenamefont {Belik},
  \citenamefont {Glazkova}, \citenamefont {Terada}, \citenamefont {Matsushita},
  \citenamefont {Sobolev}, \citenamefont {Presniakov}, \citenamefont {Tsujii},
  \citenamefont {Nimori}, \citenamefont {Takehana},\ and\ \citenamefont
  {Imanaka}}]{Belik16}%
  \BibitemOpen
  \bibfield  {author} {\bibinfo {author} {\bibfnamefont {A.~A.}\ \bibnamefont
  {Belik}}, \bibinfo {author} {\bibfnamefont {Y.~S.}\ \bibnamefont {Glazkova}},
  \bibinfo {author} {\bibfnamefont {N.}~\bibnamefont {Terada}}, \bibinfo
  {author} {\bibfnamefont {Y.}~\bibnamefont {Matsushita}}, \bibinfo {author}
  {\bibfnamefont {A.~V.}\ \bibnamefont {Sobolev}}, \bibinfo {author}
  {\bibfnamefont {I.~A.}\ \bibnamefont {Presniakov}}, \bibinfo {author}
  {\bibfnamefont {N.}~\bibnamefont {Tsujii}}, \bibinfo {author} {\bibfnamefont
  {S.}~\bibnamefont {Nimori}}, \bibinfo {author} {\bibfnamefont
  {K.}~\bibnamefont {Takehana}}, \ and\ \bibinfo {author} {\bibfnamefont
  {Y.}~\bibnamefont {Imanaka}},\ }\href {\doibase
  10.1021/acs.inorgchem.6b00774} {\bibfield  {journal} {\bibinfo  {journal}
  {Inorganic Chemistry}\ }\textbf {\bibinfo {volume} {55}},\ \bibinfo {pages}
  {6169} (\bibinfo {year} {2016})}\BibitemShut {NoStop}%
\bibitem [{\citenamefont {Verseils}\ \emph {et~al.}(2017)\citenamefont
  {Verseils}, \citenamefont {Mezzadri}, \citenamefont {Delmonte}, \citenamefont
  {Baptiste}, \citenamefont {Klein}, \citenamefont {Shcheka}, \citenamefont
  {Chapon}, \citenamefont {Hansen}, \citenamefont {Gilioli},\ and\
  \citenamefont {Gauzzi}}]{Verseils17}%
  \BibitemOpen
  \bibfield  {author} {\bibinfo {author} {\bibfnamefont {M.}~\bibnamefont
  {Verseils}}, \bibinfo {author} {\bibfnamefont {F.}~\bibnamefont {Mezzadri}},
  \bibinfo {author} {\bibfnamefont {D.}~\bibnamefont {Delmonte}}, \bibinfo
  {author} {\bibfnamefont {B.}~\bibnamefont {Baptiste}}, \bibinfo {author}
  {\bibfnamefont {Y.}~\bibnamefont {Klein}}, \bibinfo {author} {\bibfnamefont
  {S.}~\bibnamefont {Shcheka}}, \bibinfo {author} {\bibfnamefont {L.~C.}\
  \bibnamefont {Chapon}}, \bibinfo {author} {\bibfnamefont {T.}~\bibnamefont
  {Hansen}}, \bibinfo {author} {\bibfnamefont {E.}~\bibnamefont {Gilioli}}, \
  and\ \bibinfo {author} {\bibfnamefont {A.}~\bibnamefont {Gauzzi}},\ }\href
  {\doibase 10.1103/PhysRevMaterials.1.064407} {\bibfield  {journal} {\bibinfo
  {journal} {Phys. Rev. Materials}\ }\textbf {\bibinfo {volume} {1}},\ \bibinfo
  {pages} {064407} (\bibinfo {year} {2017})}\BibitemShut {NoStop}%
\bibitem [{\citenamefont {Imamura}\ \emph {et~al.}(2008)\citenamefont
  {Imamura}, \citenamefont {Karppinen}, \citenamefont {Motohashi},
  \citenamefont {Fu}, \citenamefont {Itoh},\ and\ \citenamefont
  {Yamauchi}}]{Imamura08}%
  \BibitemOpen
  \bibfield  {author} {\bibinfo {author} {\bibfnamefont {N.}~\bibnamefont
  {Imamura}}, \bibinfo {author} {\bibfnamefont {M.}~\bibnamefont {Karppinen}},
  \bibinfo {author} {\bibfnamefont {T.}~\bibnamefont {Motohashi}}, \bibinfo
  {author} {\bibfnamefont {D.}~\bibnamefont {Fu}}, \bibinfo {author}
  {\bibfnamefont {M.}~\bibnamefont {Itoh}}, \ and\ \bibinfo {author}
  {\bibfnamefont {H.}~\bibnamefont {Yamauchi}},\ }\href@noop {} {\bibfield
  {journal} {\bibinfo  {journal} {J. Am. Chem. Soc.}\ }\textbf {\bibinfo
  {volume} {130}},\ \bibinfo {pages} {14948} (\bibinfo {year}
  {2008})}\BibitemShut {NoStop}%
\bibitem [{\citenamefont {Okamoto}\ \emph {et~al.}(2009)\citenamefont
  {Okamoto}, \citenamefont {Karppinen}, \citenamefont {Yamauchi},\ and\
  \citenamefont {Fjellvåg}}]{Okamoto09}%
  \BibitemOpen
  \bibfield  {author} {\bibinfo {author} {\bibfnamefont {H.}~\bibnamefont
  {Okamoto}}, \bibinfo {author} {\bibfnamefont {M.}~\bibnamefont {Karppinen}},
  \bibinfo {author} {\bibfnamefont {H.}~\bibnamefont {Yamauchi}}, \ and\
  \bibinfo {author} {\bibfnamefont {H.}~\bibnamefont {Fjellvåg}},\ }\href
  {\doibase https://doi.org/10.1016/j.solidstatesciences.2009.03.012}
  {\bibfield  {journal} {\bibinfo  {journal} {Solid State Sciences}\ }\textbf
  {\bibinfo {volume} {11}},\ \bibinfo {pages} {1211 } (\bibinfo {year}
  {2009})}\BibitemShut {NoStop}%
\bibitem [{\citenamefont {Murakami}\ \emph {et~al.}(1998)\citenamefont
  {Murakami}, \citenamefont {Hill}, \citenamefont {Gibbs}, \citenamefont
  {Blume}, \citenamefont {Koyama}, \citenamefont {Tanaka}, \citenamefont
  {Kawata}, \citenamefont {Arima}, \citenamefont {Tokura}, \citenamefont
  {Hirota},\ and\ \citenamefont {Endoh}}]{Murakami98}%
  \BibitemOpen
  \bibfield  {author} {\bibinfo {author} {\bibfnamefont {Y.}~\bibnamefont
  {Murakami}}, \bibinfo {author} {\bibfnamefont {J.~P.}\ \bibnamefont {Hill}},
  \bibinfo {author} {\bibfnamefont {D.}~\bibnamefont {Gibbs}}, \bibinfo
  {author} {\bibfnamefont {M.}~\bibnamefont {Blume}}, \bibinfo {author}
  {\bibfnamefont {I.}~\bibnamefont {Koyama}}, \bibinfo {author} {\bibfnamefont
  {M.}~\bibnamefont {Tanaka}}, \bibinfo {author} {\bibfnamefont
  {H.}~\bibnamefont {Kawata}}, \bibinfo {author} {\bibfnamefont
  {T.}~\bibnamefont {Arima}}, \bibinfo {author} {\bibfnamefont
  {Y.}~\bibnamefont {Tokura}}, \bibinfo {author} {\bibfnamefont
  {K.}~\bibnamefont {Hirota}}, \ and\ \bibinfo {author} {\bibfnamefont
  {Y.}~\bibnamefont {Endoh}},\ }\href {\doibase 10.1103/PhysRevLett.81.582}
  {\bibfield  {journal} {\bibinfo  {journal} {Phys. Rev. Lett.}\ }\textbf
  {\bibinfo {volume} {81}},\ \bibinfo {pages} {582} (\bibinfo {year}
  {1998})}\BibitemShut {NoStop}%
\bibitem [{\citenamefont {Matsumoto}(1970)}]{Matsumoto70}%
  \BibitemOpen
  \bibfield  {author} {\bibinfo {author} {\bibfnamefont {G.}~\bibnamefont
  {Matsumoto}},\ }\href {\doibase 10.1143/JPSJ.29.606} {\bibfield  {journal}
  {\bibinfo  {journal} {Journal of the Physical Society of Japan}\ }\textbf
  {\bibinfo {volume} {29}},\ \bibinfo {pages} {606} (\bibinfo {year}
  {1970})}\BibitemShut {NoStop}%
\bibitem [{\citenamefont {Belik}\ \emph {et~al.}(2017)\citenamefont {Belik},
  \citenamefont {Matsushita}, \citenamefont {Kumagai}, \citenamefont {Katsuya},
  \citenamefont {Tanaka}, \citenamefont {Stefanovich}, \citenamefont
  {Lazoryak}, \citenamefont {Oba},\ and\ \citenamefont {Yamaura}}]{Belik17}%
  \BibitemOpen
  \bibfield  {author} {\bibinfo {author} {\bibfnamefont {A.~A.}\ \bibnamefont
  {Belik}}, \bibinfo {author} {\bibfnamefont {Y.}~\bibnamefont {Matsushita}},
  \bibinfo {author} {\bibfnamefont {Y.}~\bibnamefont {Kumagai}}, \bibinfo
  {author} {\bibfnamefont {Y.}~\bibnamefont {Katsuya}}, \bibinfo {author}
  {\bibfnamefont {M.}~\bibnamefont {Tanaka}}, \bibinfo {author} {\bibfnamefont
  {S.~Y.}\ \bibnamefont {Stefanovich}}, \bibinfo {author} {\bibfnamefont
  {B.~I.}\ \bibnamefont {Lazoryak}}, \bibinfo {author} {\bibfnamefont
  {F.}~\bibnamefont {Oba}}, \ and\ \bibinfo {author} {\bibfnamefont
  {K.}~\bibnamefont {Yamaura}},\ }\href {\doibase
  10.1021/acs.inorgchem.7b01723} {\bibfield  {journal} {\bibinfo  {journal}
  {Inorganic Chemistry}\ }\textbf {\bibinfo {volume} {56}},\ \bibinfo {pages}
  {12272} (\bibinfo {year} {2017})}\BibitemShut {NoStop}%
\bibitem [{\citenamefont {Wollan}\ and\ \citenamefont
  {Koehler}(1955)}]{Wollan1955}%
  \BibitemOpen
  \bibfield  {author} {\bibinfo {author} {\bibfnamefont {E.~O.}\ \bibnamefont
  {Wollan}}\ and\ \bibinfo {author} {\bibfnamefont {W.~C.}\ \bibnamefont
  {Koehler}},\ }\href {\doibase 10.1103/PhysRev.100.545} {\bibfield  {journal}
  {\bibinfo  {journal} {Physical Review}\ }\textbf {\bibinfo {volume} {100}},\
  \bibinfo {pages} {545} (\bibinfo {year} {1955})}\BibitemShut {NoStop}%
\bibitem [{\citenamefont {Goodenough}(1955)}]{goodenough55}%
  \BibitemOpen
  \bibfield  {author} {\bibinfo {author} {\bibfnamefont {J.~B.}\ \bibnamefont
  {Goodenough}},\ }\href {\doibase 10.1103/PhysRev.100.564} {\bibfield
  {journal} {\bibinfo  {journal} {Phys. Rev.}\ }\textbf {\bibinfo {volume}
  {100}},\ \bibinfo {pages} {564} (\bibinfo {year} {1955})}\BibitemShut
  {NoStop}%
\bibitem [{\citenamefont {Chapon}\ \emph {et~al.}(2011)\citenamefont {Chapon},
  \citenamefont {Manuel}, \citenamefont {Radaelli}, \citenamefont {Benson},
  \citenamefont {Perrott}, \citenamefont {Ansell}, \citenamefont {Rhodes},
  \citenamefont {Raspino}, \citenamefont {Duxbury}, \citenamefont {Spill},\
  and\ \citenamefont {Norris}}]{Chapon11}%
  \BibitemOpen
  \bibfield  {author} {\bibinfo {author} {\bibfnamefont {L.~C.}\ \bibnamefont
  {Chapon}}, \bibinfo {author} {\bibfnamefont {P.}~\bibnamefont {Manuel}},
  \bibinfo {author} {\bibfnamefont {P.~G.}\ \bibnamefont {Radaelli}}, \bibinfo
  {author} {\bibfnamefont {C.}~\bibnamefont {Benson}}, \bibinfo {author}
  {\bibfnamefont {L.}~\bibnamefont {Perrott}}, \bibinfo {author} {\bibfnamefont
  {S.}~\bibnamefont {Ansell}}, \bibinfo {author} {\bibfnamefont {N.~J.}\
  \bibnamefont {Rhodes}}, \bibinfo {author} {\bibfnamefont {D.}~\bibnamefont
  {Raspino}}, \bibinfo {author} {\bibfnamefont {D.}~\bibnamefont {Duxbury}},
  \bibinfo {author} {\bibfnamefont {E.}~\bibnamefont {Spill}}, \ and\ \bibinfo
  {author} {\bibfnamefont {J.}~\bibnamefont {Norris}},\ }\href@noop {}
  {\bibfield  {journal} {\bibinfo  {journal} {Neutron News}\ }\textbf {\bibinfo
  {volume} {22}},\ \bibinfo {pages} {22} (\bibinfo {year} {2011})}\BibitemShut
  {NoStop}%
\bibitem [{\citenamefont {Rodr{\'\i}guez-Carvajal}(1993)}]{Rodriguezcarvaja93}%
  \BibitemOpen
  \bibfield  {author} {\bibinfo {author} {\bibfnamefont {J.}~\bibnamefont
  {Rodr{\'\i}guez-Carvajal}},\ }\href@noop {} {\bibfield  {journal} {\bibinfo
  {journal} {Physica B}\ }\textbf {\bibinfo {volume} {192}},\ \bibinfo {pages}
  {55} (\bibinfo {year} {1993})}\BibitemShut {NoStop}%
\bibitem [{\citenamefont {Campbell}\ \emph {et~al.}(2006)\citenamefont
  {Campbell}, \citenamefont {Stokes}, \citenamefont {Tanner},\ and\
  \citenamefont {Hatch}}]{Campbell06}%
  \BibitemOpen
  \bibfield  {author} {\bibinfo {author} {\bibfnamefont {B.~J.}\ \bibnamefont
  {Campbell}}, \bibinfo {author} {\bibfnamefont {H.~T.}\ \bibnamefont
  {Stokes}}, \bibinfo {author} {\bibfnamefont {D.~E.}\ \bibnamefont {Tanner}},
  \ and\ \bibinfo {author} {\bibfnamefont {D.~M.}\ \bibnamefont {Hatch}},\
  }\href@noop {} {\bibfield  {journal} {\bibinfo  {journal} {J. Appl.
  Crystallogr.}\ }\textbf {\bibinfo {volume} {39}},\ \bibinfo {pages} {607}
  (\bibinfo {year} {2006})}\BibitemShut {NoStop}%
\bibitem [{\citenamefont {Stokes}\ \emph {et~al.}(2007)\citenamefont {Stokes},
  \citenamefont {Hatch},\ and\ \citenamefont {Campbell}}]{Stokes07}%
  \BibitemOpen
  \bibfield  {author} {\bibinfo {author} {\bibfnamefont {H.~T.}\ \bibnamefont
  {Stokes}}, \bibinfo {author} {\bibfnamefont {D.~M.}\ \bibnamefont {Hatch}}, \
  and\ \bibinfo {author} {\bibfnamefont {B.~J.}\ \bibnamefont {Campbell}},\
  }\href {http://stokes.byu.edu/isotropy.html} {\enquote {\bibinfo {title}
  {Isotropy},}\ } (\bibinfo {year} {2007})\BibitemShut {NoStop}%
\bibitem [{\citenamefont {Kumar}\ \emph {et~al.}(2017)\citenamefont {Kumar},
  \citenamefont {Yusuf},\ and\ \citenamefont {Ritter}}]{Kumar17}%
  \BibitemOpen
  \bibfield  {author} {\bibinfo {author} {\bibfnamefont {A.}~\bibnamefont
  {Kumar}}, \bibinfo {author} {\bibfnamefont {S.~M.}\ \bibnamefont {Yusuf}}, \
  and\ \bibinfo {author} {\bibfnamefont {C.}~\bibnamefont {Ritter}},\ }\href
  {\doibase 10.1103/PhysRevB.96.014427} {\bibfield  {journal} {\bibinfo
  {journal} {Phys. Rev. B}\ }\textbf {\bibinfo {volume} {96}},\ \bibinfo
  {pages} {014427} (\bibinfo {year} {2017})}\BibitemShut {NoStop}%
\bibitem [{\citenamefont {Zhang}\ \emph {et~al.}(2018)\citenamefont {Zhang},
  \citenamefont {Terada}, \citenamefont {Johnson}, \citenamefont {Khalyavin},
  \citenamefont {Manuel}, \citenamefont {Katsuya}, \citenamefont {Tanaka},
  \citenamefont {Matsushita}, \citenamefont {Yamaura},\ and\ \citenamefont
  {Belik}}]{Zhang18}%
  \BibitemOpen
  \bibfield  {author} {\bibinfo {author} {\bibfnamefont {L.}~\bibnamefont
  {Zhang}}, \bibinfo {author} {\bibfnamefont {N.}~\bibnamefont {Terada}},
  \bibinfo {author} {\bibfnamefont {R.~D.}\ \bibnamefont {Johnson}}, \bibinfo
  {author} {\bibfnamefont {D.~D.}\ \bibnamefont {Khalyavin}}, \bibinfo {author}
  {\bibfnamefont {P.}~\bibnamefont {Manuel}}, \bibinfo {author} {\bibfnamefont
  {Y.}~\bibnamefont {Katsuya}}, \bibinfo {author} {\bibfnamefont
  {M.}~\bibnamefont {Tanaka}}, \bibinfo {author} {\bibfnamefont
  {Y.}~\bibnamefont {Matsushita}}, \bibinfo {author} {\bibfnamefont
  {K.}~\bibnamefont {Yamaura}}, \ and\ \bibinfo {author} {\bibfnamefont
  {A.~A.}\ \bibnamefont {Belik}},\ }\href {\doibase
  10.1021/acs.inorgchem.8b00479} {\bibfield  {journal} {\bibinfo  {journal}
  {Inorganic Chemistry}\ }\textbf {\bibinfo {volume} {57}},\ \bibinfo {pages}
  {5987} (\bibinfo {year} {2018})},\ \bibinfo {note} {pMID:
  29722530}\BibitemShut {NoStop}%
\bibitem [{\citenamefont {Anderson}(1950)}]{anderson50}%
  \BibitemOpen
  \bibfield  {author} {\bibinfo {author} {\bibfnamefont {P.~W.}\ \bibnamefont
  {Anderson}},\ }\href {\doibase 10.1103/PhysRev.79.350} {\bibfield  {journal}
  {\bibinfo  {journal} {Phys. Rev.}\ }\textbf {\bibinfo {volume} {79}},\
  \bibinfo {pages} {350} (\bibinfo {year} {1950})}\BibitemShut {NoStop}%
\bibitem [{\citenamefont {Kanamori}(1957)}]{kanamori57}%
  \BibitemOpen
  \bibfield  {author} {\bibinfo {author} {\bibfnamefont {J.}~\bibnamefont
  {Kanamori}},\ }\href@noop {} {\bibfield  {journal} {\bibinfo  {journal} {J.
  Theoret. Phys.}\ }\textbf {\bibinfo {volume} {17}},\ \bibinfo {pages} {197}
  (\bibinfo {year} {1957})}\BibitemShut {NoStop}%
\bibitem [{\citenamefont {Goodenough}(1958)}]{goodenough58}%
  \BibitemOpen
  \bibfield  {author} {\bibinfo {author} {\bibfnamefont {J.~B.}\ \bibnamefont
  {Goodenough}},\ }\href {\doibase
  http://dx.doi.org/10.1016/0022-3697(58)90107-0} {\bibfield  {journal}
  {\bibinfo  {journal} {Journal of Physics and Chemistry of Solids}\ }\textbf
  {\bibinfo {volume} {6}},\ \bibinfo {pages} {287 } (\bibinfo {year}
  {1958})}\BibitemShut {NoStop}%
\bibitem [{\citenamefont {Radaelli}\ \emph {et~al.}(1997)\citenamefont
  {Radaelli}, \citenamefont {Cox}, \citenamefont {Marezio},\ and\ \citenamefont
  {Cheong}}]{Radaelli97}%
  \BibitemOpen
  \bibfield  {author} {\bibinfo {author} {\bibfnamefont {P.~G.}\ \bibnamefont
  {Radaelli}}, \bibinfo {author} {\bibfnamefont {D.~E.}\ \bibnamefont {Cox}},
  \bibinfo {author} {\bibfnamefont {M.}~\bibnamefont {Marezio}}, \ and\
  \bibinfo {author} {\bibfnamefont {S.~W.}\ \bibnamefont {Cheong}},\ }\href
  {\doibase 10.1103/PhysRevB.55.3015} {\bibfield  {journal} {\bibinfo
  {journal} {Physical Review B}\ }\textbf {\bibinfo {volume} {55}},\ \bibinfo
  {pages} {3015} (\bibinfo {year} {1997})}\BibitemShut {NoStop}%
\bibitem [{\citenamefont {Tokura}(2006)}]{tokura06}%
  \BibitemOpen
  \bibfield  {author} {\bibinfo {author} {\bibfnamefont {Y.}~\bibnamefont
  {Tokura}},\ }\href {http://stacks.iop.org/0034-4885/69/i=3/a=R06} {\bibfield
  {journal} {\bibinfo  {journal} {Reports on Progress in Physics}\ }\textbf
  {\bibinfo {volume} {69}},\ \bibinfo {pages} {797} (\bibinfo {year}
  {2006})}\BibitemShut {NoStop}%
\bibitem [{\citenamefont {Johnson}\ \emph {et~al.}(2017)\citenamefont
  {Johnson}, \citenamefont {Khalyavin}, \citenamefont {Manuel}, \citenamefont
  {Radaelli}, \citenamefont {Glazkova}, \citenamefont {Terada},\ and\
  \citenamefont {Belik}}]{Johnson17}%
  \BibitemOpen
  \bibfield  {author} {\bibinfo {author} {\bibfnamefont {R.~D.}\ \bibnamefont
  {Johnson}}, \bibinfo {author} {\bibfnamefont {D.~D.}\ \bibnamefont
  {Khalyavin}}, \bibinfo {author} {\bibfnamefont {P.}~\bibnamefont {Manuel}},
  \bibinfo {author} {\bibfnamefont {P.~G.}\ \bibnamefont {Radaelli}}, \bibinfo
  {author} {\bibfnamefont {I.~S.}\ \bibnamefont {Glazkova}}, \bibinfo {author}
  {\bibfnamefont {N.}~\bibnamefont {Terada}}, \ and\ \bibinfo {author}
  {\bibfnamefont {A.~A.}\ \bibnamefont {Belik}},\ }\href {\doibase
  10.1103/PhysRevB.96.054448} {\bibfield  {journal} {\bibinfo  {journal} {Phys.
  Rev. B}\ }\textbf {\bibinfo {volume} {96}},\ \bibinfo {pages} {054448}
  (\bibinfo {year} {2017})}\BibitemShut {NoStop}%
\end{thebibliography}%

\appendix
\section{Irreducible representations for the commensurate propagation vectors}

\begin{table}[h!]
\caption{\label{TAB:irreps}Basis vectors of the irreducible representations used to describe the commensurate magnetic structures of $R$Mn$_7$O$_{12}$. Sites related by $I$ centring (not listed here) respect the propagation vector according to the phase $2\pi\mathbf{k}\cdot\mathbf{r}$. A cartesian basis is adopted, where $x || a^*$, $y || b$, and $z || c$.}
\begin{ruledtabular}
\begin{tabular}{l| c c c c }
& $\Gamma_1^+$ & $\Gamma_2^+$ & $\mathrm{Y}_1^-$ & $\mathrm{Y}_2^-$ \\
\hline
\textbf{k} & (0,0,0) & (0,0,0) & (0,1,0) & (0,1,0) \\
\hline
Mn1,                                     & [0,1,0] & [1,0,0] & - & - \\
$0,\tfrac{1}{2},0$                       &         & [0,0,1] & - & - \\
                                         & \\
Mn2,                                     & [0,1,0] & [1,0,0] & - & - \\
$\tfrac{1}{2},0,0$                       &         & [0,0,1] & - & - \\
                                         & \\
Mn3,                                     & [0,1,0] & [1,0,0] & - & - \\
$\tfrac{1}{2},\tfrac{1}{2},0$            &         & [0,0,1] & - & - \\
                                         & \\
Mn4,                                     & [1,0,0] & [1,0,0] & [1,0,0] & [1,0,0] \\
$\tfrac{1}{4},\tfrac{1}{4},\tfrac{1}{4}$ & [0,1,0] & [0,1,0] & [0,1,0] & [0,1,0] \\
                                         & [0,0,1] & [0,0,1] & [0,0,1] & [0,0,1] \\
                                         & \\
Mn4,                                     & [$\bar{1}$,0,0] & [1,0,0] & [1,0,0] & [$\bar{1}$,0,0] \\
$\tfrac{1}{4},\tfrac{3}{4},\tfrac{1}{4}$ & [0,1,0] & [0,$\bar{1}$,0] & [0,$\bar{1}$,0] & [0,1,0] \\
                                         & [0,0,$\bar{1}$] & [0,0,1] & [0,0,1] & [0,0,$\bar{1}$] \\
                                         & \\
Mn5,                                     & [1,0,0] & [1,0,0] & [1,0,0] & [1,0,0] \\
$\tfrac{3}{4},\tfrac{1}{4},\tfrac{1}{4}$ & [0,1,0] & [0,1,0] & [0,1,0] & [0,1,0] \\
                                         & [0,0,1] & [0,0,1] & [0,0,1] & [0,0,1] \\
                                         & \\
Mn5,                                     & [$\bar{1}$,0,0] & [1,0,0] & [1,0,0] & [$\bar{1}$,0,0] \\
$\tfrac{3}{4},\tfrac{3}{4},\tfrac{1}{4}$ & [0,1,0] & [0,$\bar{1}$,0] & [0,$\bar{1}$,0] & [0,1,0] \\
                                         & [0,0,$\bar{1}$] & [0,0,1] & [0,0,1] & [0,0,$\bar{1}$] \\
\end{tabular}
\end{ruledtabular}
\end{table}

\end{document}